\documentclass[superscriptaddress,showkeys,secnumarabic,twocolumn]{revtex4}%
\usepackage{amsfonts}
\usepackage{amsmath}
\usepackage{amssymb}
\usepackage{graphicx}%
\providecommand{\U}[1]{\protect \rule{.1in}{.1in}}
\newlength {\defaultparindent }
\newenvironment {annotation Text}{}{}

\newcommand \beginfigure{\begin{figure}[p]}

\begin{document}
\title{Reciprocity in quantum, electromagnetic and other wave scattering}
\author{L. De\'{a}k}
\email{deak@rmki.kfki.hu}
\affiliation{KFKI Research Institute for Particle and Nuclear Physics, P.O.B. 49, H-1525
Budapest, Hungary}
\author{T. F\"{u}l\"{o}p}
\email{fulopt@rmki.kfki.hu}
\affiliation{KFKI Research Institute for Particle and Nuclear Physics, P.O.B. 49, H-1525
Budapest, Hungary}

\begin{abstract}
The reciprocity principle is that, when an emitted wave gets scattered on an
object, the scattering transition amplitude does not change if we interchange
the source and the detector -- in other words, if incoming waves are
interchanged with appropriate outgoing ones. Reciprocity is sometimes confused
with time reversal invariance, or with invariance under the rotation that
interchanges the location of the source and the location of the detector.
Actually, reciprocity covers the former as a special case, and is
fundamentally different from -- but can be usefully combined with -- the latter.
Reciprocity can be proved as a theorem in many situations and is found
violated in other cases. The paper presents a general treatment of
reciprocity, discusses important examples, shows applications in the field of
photon (M\"ossbauer) scattering, and establishes a fruitful connection with a
recently developing area of mathematics. 

\end{abstract}
\maketitle

\preprint{HEP/123-qed}

\section{Introduction\label{Introduction}}

Symmetry is a key notion in physics. In the field of natural sciences,
symmetry can be interpreted as a concept of balance or patterned
self-similarity \cite{Weyl1982}, regarding to concrete as well as abstract
objects like a physical system itself or, for example, a theoretical model,
respectively \cite{Klaus2005}. According to the Oxford online dictionary,
symmetry is a law or operation where a physical property or process has an
equivalence in two or more directions, or events/actions are balanced/equal in
some way. A synonym to the term symmetry is invariance, expressing the fact
that a special operation (called symmetry transformation), which could be a
change of some physical parameters -- like a geometric transformation, a change
of polarization, parity, charge, the arrow of time, etc. -- does not change
some particular property of the system. Depending on the studied system,
symmetry may have various measures and operational definitions. In the
theoretical model of quantum mechanics, symmetry transformations were given by
\textcite{wigner} as general operators preserving the modulus of scalar
products of the vectors of the
Hilbert space representing the physical states. Further, the symmetry theorem
of \textcite{wigner} states that any symmetry transformation can be
represented by either a unitary linear or an isometric conjugate linear
(usually called antiunitary) operator.

\begin{figure}[p]
\resizebox{.6\columnwidth}{!}{\includegraphics[clip]{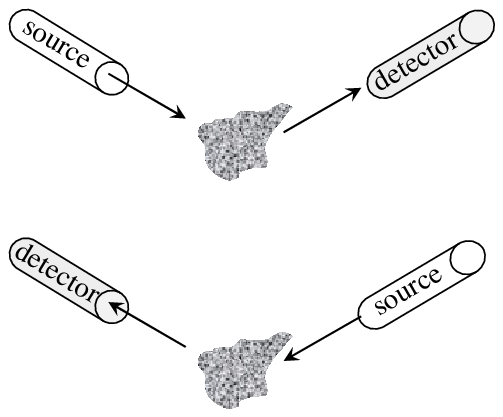}}
\caption{Scattering
experiment in the original (upper) and the reversed (lower) arrangements. The
reciprocity principle means invariance of the scattering transition amplitude
under the reversal of the source and the detector.}%
\label{fig1}%
\end{figure}

A large class of physical processes can be described in the framework of
scattering theory \cite{SCHIFF,MessiahII}, which is a theoretical sub-model
inside quantum mechanics, but can be interpreted in classical electrodynamics
and in the classical mechanics of elastic waves as well. As documented in the
literature since long ago, physical intuition suggests that, when reversing
the position of source and detector in a wave scattering experiment\ (see
Fig.~\ref{fig1})%
, the observed signal will not change. This condition, called the reciprocity
principle, is indeed fulfilled in most cases. Nevertheless, it does not follow
directly from first principles, therefore, it is not necessarily fulfilled.

The physical term `reciprocity' appeared already in the
19$^{\mathrm{th}}$ century. The first reciprocity equations related
reflection and transmission of light at an interface between lossless
optical media, and were derived by \textcite{Stokes1849}. The principle
of reciprocity was later generalized for more complex scattering systems
in the field of electromagnetic waves \cite{Helmholtz1866,Lorentz1905},
sound waves \cite{Strutt1877}, electric circuits \cite{Carson1924}, and
radio communication \cite{Carson1930}, as well as in quantum mechanical
scattering problems \cite{Bilhorn1964}. The reciprocity related
publications cover the whole 20$^{\mathrm{th}}$ century, as it is
summarized in the review paper of \textcite{Potton2004}. Reciprocity can
be proven for various scattering problems with certain limits of
validity \cite{deHoop,Saxon1955,Carminati2000,Hillion1978}.

The number of reciprocity related publications grew intensively in the last
decade as well, and all the approaches that appeared earlier were subject to
further developments. Among others, the reciprocity relations of
\textcite{Stokes1849} were generalized to multilayers \cite{Vigoureux2000} and
to absorptive multilayers \cite{Gigli2001,Andre2009}. The applications in
electrodynamics for electric circuits and antennas were studied by
\textcite{Sevegi2010}, and the reciprocity theorem of classical
electrodynamics in case of material media containing linearly polarizable and
linearly magnetizable substances was formulated by \textcite{Mansuripur2011}.
Nonreciprocal devices (circulators and isolators) with on-chip integration
possibility were recently suggested by \textcite{Kamal2011}. In parallel,
reciprocity admits applications in particle scattering, acoustics, seismology,
and the solution of inverse problems as well \cite{Potton2004}.

Reciprocity was considered for nonlocal electrodynamical
\cite{Xie2009,Xie2009b} and nonlocal quantum mechanical systems \cite{Xie2008}%
. In a recent publication of \textcite{Leung2010}, the aspect of gauge
invariance was discussed from the point of view of quantum mechanical
interpretation of reciprocity, and new gauge invariant formulations of
reciprocity were suggested and analyzed.

What is reciprocity? In many works it is simply related to time reversal
symmetry \cite{Mytnichenko2005,Wurmser1996} as was done in the well-known
reciprocity theorem of \textcite{LandauIII}. The optical reciprocity theorem,
however, revealed that absorption, which violates time reversal invariance,
conserves reciprocity in polarization independent cases \cite{BornWolf}, which
observation was also expressed in scattering theory \cite{Bilhorn1964}. In
parallel, according to the original reciprocity principle, namely, invariance
under the interchange of source and detector, one could have the impression
that reciprocity is identical to a rotation by $180^{\circ}$. However, this
latter interpretation also proves false since there exist scatterers with no
$180^{\circ}$ rotational symmetry but fulfilling the reciprocity principle.

The currently typically used condition of reciprocity in linear systems
is the self-transpose (also called complex symmetric) property of the
matrix of the scattering potential, of the index of refraction, of the
dielectric/magnetic permeability tensors, or of the Green's function
\cite{Xie2009,Xie2009b}. This condition, however, depends on the frame,
on the polarization basis chosen. Indeed, applying a unitary basis
transformation, the self-transpose property of the matrix is not
conserved, as it can be demonstrated on the case of a Hermitian matrix,
which is not self-transpose in general, but can be diagonalized --
hence, the self-transpose form is obtained by an appropriate unitary
transformation. In the light of this observation, the physical content
of reciprocity seems to be unclear. Our main task is to give a proper
frame-independent description of reciprocity, extending the excellent
early work of \textcite{Bilhorn1964}.

The nonreciprocal properties of systems are even less understood.
Magneto-optical systems are typically cited as nonreciprocal media
\cite{Potton2004,Kamal2011}, but detailed analyses of the reasons of
reciprocity violation have not been given.

Reciprocity violation can be obtained in case of magneto-optical gyrotropy
\cite{Potton2004}, which is a well-known property of the M\"{o}ssbauer medium
\cite{Blume68}. At M\"{o}ssbauer resonances, the ratio of the
time-inversion-violating to normal potentials is typically of the order of one
thousand! The resonant M\"{o}ssbauer medium is absorptive and gyrotropic and,
accordingly, for well-defined geometrical situations, significant reciprocity
violation is expected.

The content of reciprocity is the same for any type of classical wave as well
as for quantum mechanics. For definiteness, we discuss reciprocity in the
quantum mechanical framework, in scattering theory that corresponds to the
Schr\"{o}dinger equation. Note that any classical wave equation can be
rewritten in the form of a Schr\"{o}dinger equation [as done, for example, by
\textcite[p347]{Richtmyer}, \textcite[Ch.~I \S 1.]{AkhBer},
\textcite{SakTak40}, and \textcite{FesVil}], and, under this correspondence,
what is probability density in the quantum context, is energy density in case
of classical waves. Especially, the case of two-component wave function, which
in quantum mechanics describes a 1/2 spin particle, \textit{e.g.,} a neutron,
is equally able to represent the two transversal polarization degrees of
freedom of photon so the scattering of slow neutrons and that of photons admit
a common formalism \cite{Lax51}, \cite{Deak01}.

In Part~\ref{raa}, we present the general formalism of reciprocity.
Part~\ref{rab} investigates the case of two spin/polarization degrees of
freedom in detail, and the results are illustrated and applied on examples
related to the area of M\"{o}ssbauer scattering in Part~\ref{rac}.

Our discussion analyzes the relationship of reciprocity to time reversal
invariance and to rotational invariance, studies a form of quasireciprocity
and the specialties emerging in the Born approximation, and uncovers a link
to a recently expanding area of mathematics the results of which assist
physics in identifying and finding systems with the reciprocity property.

\section{The general formulation of reciprocity\label{raa}}

To formulate reciprocity, let us first revisit two topics involved, scattering
theory and antiunitary operators, briefly summarizing the ingredients utilized
in what follows.

\subsection{Notations: Scattering theory\label{rau}}

Concerning scattering theory, we use notations, conventions, and standard
results from \textcite{SCHIFF} and \textcite{MessiahII}; see also
\textcite{Galindo-Pascual} and \textcite{Reed-SimonIII}, for example, for
technical details.

Let $H_{0}$ be a self-adjoint Hamiltonian, which, for simplicity, will be
called a free Hamiltonian although it need not really describe a free
quantum/wave propagation -- for example, neutrons emitted by a source may
travel through a guiding magnetic field. Furthermore, let a potential $V$
describe a scatterer. $V$ is not assumed to be self-adjoint, which allows
absorption effects to be incorporated.

With the stationary Green's operators
\begin{equation}
G^{\pm}_{{\hskip-.02em\scriptscriptstyle E}}:=\left( E-H\pm i\epsilon \right)
^{-1} \label{rcw}%
\end{equation}
($\epsilon \searrow0$ understood), if $u$ is an eigenstate of $H_{0}$ with a
real eigenvalue $E$ then, under suitable conditions on $V$, the states
introduced as
\begin{align}
\chi^{\pm} & :=u+G^{\pm}_{{\hskip-.02em\scriptscriptstyle E}}Vu,\label{Lip1}\\
\chi^{T\pm} & :=u+ {G^{\mp}_{{\hskip-.02em\scriptscriptstyle E}}} ^{\dag
}V^{\dag}u\label{Lip2}%
\end{align}
prove to be such $E$ eigenvalued eigenstates of $H=H_{0}+V$ and its adjoint
$H^{\dag}$, respectively, that $u$ is their asymptotically incoming (`$+$'
sign) or outgoing (`$-$' sign) part.

For any two eigenstates $u_{\alpha}$, $u_{\beta}$ of $H_{0}$ with eigenvalue
$E_{\alpha}=E_{\beta}$, the $u_{\alpha}\rightarrow u_{\beta}$ elastic
scattering transition amplitude reads and satisfies
\begin{equation}
\left \langle \beta \left \vert T\right \vert \alpha \right \rangle := \left(
u_{\beta}^{},V\chi_{\alpha}^{+}\right) = \big (\chi_{\beta}^{T-},Vu_{\alpha
}^{}\big ). \label{trans_ampl}%
\end{equation}

In case the scatterer can be divided into two sub-scatterers, $V=V_{1}+V_{2}$,
the transition amplitude can also be given as a sum as
\begin{equation}
\left \langle \beta \left \vert T\right \vert \alpha \right \rangle =\left \langle
\beta \left \vert T_{1}\right \vert \alpha \right \rangle +\left( \chi_{1\beta
}^{T-},V_{2}\chi_{\alpha}^{+}\right) ,\label{twopot}%
\end{equation}
where $\left \langle \beta \left \vert T_{1}\right \vert \alpha \right \rangle $ is
the transition amplitude of scattering on $V_{1}$ alone, and $\chi_{1\beta
}^{T-}$ also corresponds to $V_{1}$ only. Naturally, the role of $V_{1}$ and
$V_{2}$ can be interchanged here.

As for approximations -- to which one is forced to resort in many
applications, -- the (1$^{{\mathrm{st}}}$) Born approximation is when the
scattering solutions are replaced by the corresponding free ones,
\begin{equation}
\chi^{+}\approx u,\qquad \chi^{T-}\approx u,\label{rcl}%
\end{equation}
and thus (\ref{trans_ampl}) is approximated as
\begin{equation}
\left \langle \beta \left \vert T\right \vert \alpha \right \rangle \approx \left(
u_{\beta},Vu_{\alpha}\right) ,\label{rdo}%
\end{equation}
and (\ref{twopot}) as
\begin{equation}
\left \langle \beta \left \vert T\right \vert \alpha \right \rangle \approx \left(
u_{\beta},V_{1}u_{\alpha}\right) + \left( u_{\beta},V_{2}u_{\alpha}\right)
.\label{rdp}%
\end{equation}

We can see that scattering formulae get considerably simplified in the Born approximation.

\subsection{Notations: Antiunitary operators\label{rat}}

The key notion behind reciprocity is the notion of antiunitary (also called
conjugate unitary) operators. On a separable complex Hilbert space
${\mathcal{H}}$, an operator $U:{\mathcal{H}}\to{\mathcal{H}}$ is called
unitary if it is isometric,
\begin{equation}
\|U\psi \|=\| \psi \|,\qquad \psi \in{\mathcal{H}},\label{raw}%
\end{equation}
and linear,
\begin{align}
U\left( \lambda_{1}^{}\psi_{1}^{}+\lambda_{2}^{}\psi_{2}^{}\right)   &
=\lambda_{1}^{}U\psi_{1}^{}+\lambda_{2}^{}U\psi_{2}^{},\nonumber \\
\lambda_{1}^{},\lambda_{2}^{} \in{\mathbb{C}},  &
\phantom{\rule{1.2em}{2.8ex}} \psi_{1}^{},\psi_{2}^{} \in{\mathcal{H}%
}.\label{rav}%
\end{align}
For antiunitary operators $K:{\mathcal{H}}\to{\mathcal{H}}$, isometry remains
valid,
\begin{equation}
\|K\psi \|=\| \psi \|,\qquad \psi \in{\mathcal{H}},\label{rax}%
\end{equation}
while linearity is replaced by antilinarity (conjugate linearity),
\begin{align}
K\left( \lambda_{1}^{}\psi_{1}^{}+\lambda_{2}^{}\psi_{2}^{}\right)   &
=\lambda_{1}^{*}K\psi_{1}^{}+\lambda_{2}^{*}K\psi_{2}^{},\nonumber \\
\lambda_{1}^{},\lambda_{2}^{} \in{\mathbb{C}},  &
\phantom{\rule{1.2em}{2.8ex}} \psi_{1}^{},\psi_{2}^{} \in{\mathcal{H}%
},\label{ray}%
\end{align}
where ${}^{*}$ denotes complex conjugation. In both cases, isometry implies
the existence of the inverse operator, and the inverse proves to coincide with
the adjoint,
\begin{equation}
U^{-1}=U^{\dag},\qquad K^{-1}=K^{\dag},\label{raz}%
\end{equation}
which are again unitary and antiunitary, respectively. With the aid of the
so-called polarization identity, from the isometric property one finds, for
any $\psi_{1},\psi_{2}\in{\mathcal{H}}$,
\begin{equation}
\left( U\psi_{1},U\psi_{2}\right) =\left( \psi_{1},\psi_{2}\right)
\, \label{rba}%
\end{equation}
and
\begin{equation}
\left( K\psi_{1},K\psi_{2}\right) =\left( \psi_{2},\psi_{1}\right)
,\label{rbb}%
\end{equation}
respectively. It is this scalar product swapping property (\ref{rbb}) that
will be shown below to be the key point why reciprocity is connected to
\textit{antiunitary} operators.

The most frequently treated antiunitary operators are the involutive ones, and
are called conjugations. Namely, an antiunitary operator $C$ is a conjugation
if it is involutive,
\begin{equation}
C^{2}=I,\label{rbc}%
\end{equation}
with $I$ denoting the identity operator of the Hilbert space ${\mathcal{H}}$.
Similarly defined are the anticonjugations, those antiunitary operators whose
square is $-I$ rather than $I$ (antiinvolutions). The most well-known example
for a conjugation operator is the standard complex conjugation
\begin{equation}
J\psi=\psi^{*}\label{rbi}%
\end{equation}
of complex functions $\psi$ in an ${\mathcal{L}}^{2}$ Hilbert space.
Conjugations possess various nice properties. For example, any conjugation
admits an orthonormal eigenbasis $e_{1},e_{2},\ldots$ in ${\mathcal{H}}$ with
unit eigenvalues, $c_{1}=c_{2}=\cdots=1$; and, conversely, any orthonormal
basis defines a conjugation by being its eigenbasis with eigenvalues $1$.

Antiunitary operators are the same in number as unitary operators, in the
standard sense that they can be brought into one-to-one correspondence.
Indeed, choosing an arbitrary antiunitary operator -- for later purposes, let
it actually be a conjugation $C$ --  any antiunitary $K$ can be written in the
form
\begin{equation}
K=UC,\label{rbd}%
\end{equation}
where $U$ is unitary. In fact,
\begin{equation}
U:=KC^{-1}\label{r}%
\end{equation}
is a product of two isometric and antilinear operators, thus being isometric
and linear, \textit{i.e.,} unitary. Conversely, the multiplication of any
unitary $U$ with $C$ is similarly found to give an antiunitary $K:=UC$.

In quantum mechanics, any symmetry can be given via either a unitary or an
antiunitary operator. In practice, antiunitary cases are much less frequently
encountered than unitary ones. Two well-known antiunitary symmetries are
charge conjugation and time reversal; the former being a conjugation (in the
above sense, having $C^{2}=I$) and the latter being either a conjugation or an
anticonjugation, depending on particle number and spin.

The subsequent considerations involve not only conjugations or
anticonjugations but arbitrary antiunitary operators.

\subsection{The reciprocity condition and its consequences\label{rbx}}

In a scattering situation as described before, let us assume that an
antiunitary operator $K$ commutes with the free Hamiltonian,
\begin{equation}
KH_{0}K^{-1}=H_{0},\label{free-com}%
\end{equation}
and also that it connects the potential $V$ with its adjoint as
\begin{equation}
KVK^{-1}=V^{\dag}.\label{CdefV}%
\end{equation}
Then, for the full Hamiltonian $H=H_{0}+V$, we have
\begin{equation}
KHK^{-1}=H^{\dag}.\label{CdefH}%
\end{equation}
Eqs.~(\ref{free-com})--(\ref{CdefV}) can be called the \textit{reciprocity
conditions}, and $K$ a \textit{reciprocity operator} for the system -- the
reason for these names will be clear soon.

A consequence of (\ref{free-com}) is that, if $u$ is an eigenstate of $H_{0}$
with real eigenvalue $E$, then $Ku$ is also its eigenstate, and possesses the
same eigenvalue $E$.

In parallel, for any real $E$, (\ref{CdefH}) implies
\begin{align}
K\left( E-H\pm i\epsilon \right) K^{-1} & =\left( E-H^{\dag}\mp i\epsilon
\right) ,\label{raj}\\
K\left( E-H\pm i\epsilon \right) ^{-1}K^{-1} & =\left( E-H^{\dag}\mp
i\epsilon \right) ^{-1},\label{rak}%
\end{align}
the latter following from the former. Eq.~(\ref{rak}) formulates the
\textit{reciprocity theorem for the Green's operator} (\ref{rcw}):
\begin{equation}
KG^{\pm}_{{\hskip-.02em\scriptscriptstyle E}}K^{-1}={G^{\pm}%
_{{\hskip-.02em\scriptscriptstyle E}}}^{\dag}.\label{rej}%
\end{equation}

Next, we study the consequences of the reciprocity property on the scattering
quantities. Considering the $u_{\alpha}\rightarrow u_{\beta}$ elastic
scattering transition amplitude, let us introduce the notations
\begin{equation}
u_{\overline{\alpha}}:=Ku_{\alpha},\qquad u_{\overline{\beta}}:=Ku_{\beta
}.\label{rah}%
\end{equation}
$K$ commutes with $H_{0}$ and we are treating an elastic process so
\begin{equation}
E_{\alpha}=E_{\overline{\alpha}}=E_{\beta}=E_{\overline{\beta}}=:E.\label{rbf}%
\end{equation}
Applying (\ref{CdefV}) and (\ref{rej}) on the definitions (\ref{Lip1}%
)--(\ref{Lip2}), one finds
\begin{equation}
K\chi_{\alpha}^{\pm}=\chi_{\overline{\alpha}}^{T\mp},\qquad K\chi_{\beta}%
^{\pm}=\chi_{\overline{\beta}}^{T\mp}.\label{rai}%
\end{equation}
In words, $K$ maps a scattering process to a ``reversed'' one. Hence, for the
transition amplitude, we obtain
\begin{align}
\left \langle \beta \left \vert T\right \vert \alpha \right \rangle  & =\left(
u_{\beta}^{},V\chi_{\alpha}^{+}\right) \nonumber \\
& =\left( KV\chi_{\alpha}^{+},Ku_{\beta}^{}\right) =\left( V^{\dagger}%
K\chi_{\alpha}^{+},Ku_{\beta}^{}\right) \nonumber \\
& =\left( K\chi_{\alpha}^{+},VKu_{\beta}^{}\right) =\left( \chi_{\overline
{\alpha}}^{T-},Vu_{\overline{\beta}}^{}\right) \nonumber \\
& =\left \langle \overline{\alpha}\left \vert T\right \vert \overline{\beta
}\right \rangle ,\label{rbg}%
\end{align}
where (\ref{trans_ampl}), (\ref{rbb}), (\ref{CdefV}), (\ref{rai}),
(\ref{rah}), and again (\ref{trans_ampl}) have been utilized, in turn.

This result, (\ref{rbg}), is the \textit{reciprocity theorem for the
transition amplitude}. Why it is in fact the manifestation of the reciprocity
principle -- which has been explained in the Introduction -- is clear from that
it relates a scattering process to a ``reversed'' one. As anticipated in
Sect.~\ref{rat}, it is the left-right interchanging property (\ref{rbb}) that
makes this ``reciprocal'' relation possible, which explains why an antiunitary
transformation is the heart of reciprocity.

This immediately indicates that \textit{reciprocity is not the same as
rotational invariance} under some rotation: Rotations always act on Hilbert
space vectors as unitary, not antiunitary, operators. (More on rotations
vs.\ reciprocity follows in Sect.~\ref{Rotavsreci}.)

Let us observe that the adjoint of (\ref{CdefV}) reads
\begin{equation}
KV^{\dag}K^{-1}=V,\label{rcb}%
\end{equation}
which can also be rearranged as
\begin{equation}
V^{\dag}=K^{-1}VK.\label{rcc}%
\end{equation}
Comparing this (\ref{rcc}) with (\ref{CdefV}) shows that $K^{-1}$ is also a
reciprocity operator. In parallel, (\ref{rcb}) helps us to derive
\begin{align}
K^{2}VK^{-2} & =K\left( KVK^{-1}\right) K^{-1} =KV^{\dag}K^{-1}=V.\label{rca}%
\end{align}
This says that $V$ and $K^{2}$ commute, or, in other words, that $K^{2}$ is a
symmetry of the system. Naturally, then $K^{4},K^{6},\ldots$ and
$K^{-2},K^{-4},\ldots$ each also commute with $V$. As a consequence, all the
antiunitary operators $K^{3},K^{5},\ldots$ and $K^{-1},K^{-3},K^{-5},\ldots$
are also reciprocity operators of the system. This allows further
(\ref{rbg})-type formulae, in which $u_{\alpha}$ is related not to
$u_{\overline{\alpha}}=Ku_{\alpha}$ but to $u_{{\overline{\alpha}}^{(3)}%
}:=K^{3}u_{\alpha}$, etc.

\subsection{Reciprocal partner systems\label{rer}}

If, more generally, $K$ does not fulfill (\ref{CdefV}) but provides a
connection with another scattering problem with potential $\overline{V}$,
\begin{equation}
KVK^{-1}=\overline{V}^{\dag},\label{ral}%
\end{equation}
then a completely analogous calculation provides
\begin{equation}
\left \langle \beta \left \vert T\right \vert \alpha \right \rangle =\left(
\overline{\chi}_{\overline{\alpha}}^{T-},\overline{V}u_{\overline{\beta}}%
^{}\right) =\big \langle \overline{\alpha}\big \vert \overline{T}%
\big \vert \overline{\beta}\big \rangle ,\label{ram}%
\end{equation}
where $\overline{\chi}$ and $\overline{T}$ correspond to $\overline{V}$,
according to the sense. In such a case we can call the system with
$\overline{H}$ the \textit{reciprocal partner} of the system with $H$.

The relationship between a system and its reciprocal partner is a duality type
one, \textit{i.e.,} the reciprocal partner of a reciprocal partner is the
original system. Indeed, taking the adjoint of (\ref{ral}) leads to
\begin{equation}
K\overline{V}K^{-1}=V^{\dag}.\label{rdl}%
\end{equation}

Note that, for any $K$, the operator defined as
\begin{equation}
\overline{V}:=\left( KVK^{-1}\right) ^{\dag}=KV^{\dag}K^{-1}\label{rbh}%
\end{equation}
trivially automatically fulfills (\ref{ral}). The nontrivial question here is
whether this definition provides just a mere abstract Hilbert space operator
or a physically reasonable scattering potential, for which scattering theory
also holds.

In quantum mechanics, if two Hamiltonians $H$, $\breve{H}$ are connected by a
unitary transformation
\begin{equation}
\breve{H}=U_{\phi}^{\phantom {|}}HU_{\phi}^{-1}\label{rcp}%
\end{equation}
with a unitary multiplying operator $U_{\phi}=e^{i\phi(t,{\mathbf{r}})}$
acting on wave functions as
\begin{equation}
\breve{\psi}(t,{\mathbf{r}})=e^{i\phi(t,{\mathbf{r}})}\psi(t,{\mathbf{r}%
}),\label{rcr}%
\end{equation}
then $H$ and $\breve{H}$ describe the same physical system with the wave
functions $\psi$ mapped to $\breve{\psi}$ according to (\ref{rcr}) and any
physical quantity operator $O$ mapped to $\breve{O}$ via $\, \breve{O}=U_{\phi
}^{\phantom {|}}OU_{\phi}^{-1}\,$ . Eq.~(\ref{rcr}) is the most general
transformation freedom under the requirement $|\breve{\psi}|^{2}=|\psi|^{2}$,
the preservation of the position representation. For a charged quantum
particle in an electromagnetic field, such a so-called gauge transformation of
quantities is accompanied by the gauge transformation
\begin{equation}
\breve{V}=V-(\hbar/q)\partial_{t}\phi,\qquad \breve{{\mathbf{A}}}={\mathbf{A}%
}+(\hbar/q)\nabla \phi \label{rct}%
\end{equation}
of the electromagnetic four-potential ($q$ being the charge and $\hbar$ the
reduced Planck constant).

If two such gauge equivalent Hamiltonians are connected by an antiunitary $K$
as
\begin{equation}
KH_{0}K^{-1}=\breve{H}_{0}\,,\quad KVK^{-1}=\breve{V}^{\dag}\,,\quad
KHK^{-1}=\breve{H}^{\dag}\label{rcu}%
\end{equation}
then this is actually a reciprocity property of one and the same physical
system, and $K$ can still be called a reciprocity operator of this system. It
is only that two -- different but gauge equivalent -- representations of this
system are appearing in the formulae. The gauge aspect has been emphasized by
\textcite{Leung2010}. If there is no such gauge connection between a $H$ and a
$\overline{H}$ then they do represent two different physical systems that are
in a reciprocity relationship.

\subsection{Time reversal is a special case\label{rag}}

If a system possesses not only the reciprocity properties
(\ref{free-com})--(\ref{CdefV}) but also $V=V^{\dag}$, $H=H^{\dag}$ then
$KH=HK$ so $K$ is actually a symmetry of the system. Conversely, if
$H=H^{\dag}$ and $K$ is a symmetry for both $H_{0}$ and $H$, \textit{i.e.,}
$KH_{0}=H_{0}K$ and $KH=HK$, then $K$ is a reciprocity operator.

A seminal special case is when $K$ is the time reversal operator. That is, if
there is no absorption and time reversal is a symmetry then the reciprocity
theorem (\ref{rbg}) holds and coincides with what is usually called the
microreversibility property of the scattering amplitude
\cite{SCHIFF,MessiahII}.

Nevertheless, let us observe that, even if time reversal is not a
symmetry -- \textit{e.g.,} if absorption is present --  still we can have a
reciprocity theorem offered by the time reversal operator, as long as it
fulfills the reciprocity conditions (\ref{free-com})--(\ref{CdefV}). In such
cases, time reversal ensures a property which cannot be called
microreversibility any more but is still a reciprocity property.

Therefore, \textit{reciprocity is not the same as time reversal invariance:}
\newline-- time reversal is by far not the only possible reciprocity operator,
and \newline-- it can be a reciprocity operator (source of a reciprocity
theorem) irrespective of whether it is a symmetry of the system.

Time reversal is revisited in Sect.~\ref{raf}.

\subsection{Connection with recent mathematical results\label{rdd}}

Similarly to that a Hamiltonian may, but not necessarily does, admit a
symmetry, we may ask how frequently it occurs that the reciprocity conditions
(\ref{free-com})--(\ref{CdefV}) are satisfied by some appropriate antiunitary
operator $K$. Typically it is easier to check $K$ against the free Hamiltonian
and the less easy task is to investigate the validity of (\ref{CdefV}). To get
closer to the answer to the latter question, let us choose an auxiliary
conjugation $C$ arbitrarily, with the aid of which we describe the various
possible $K$'s by the various possible $U$'s via (\ref{rbd}).

This way, we can rewrite (\ref{CdefV}) as
\begin{align}
V=\left( KVK^{-1}\right) ^{\dag}  &  = \left( UCVC^{-1}U^{-1}\right) ^{\dag
}\nonumber \\
&  = U\left( CVC^{-1}\right) ^{\dag}U^{-1}.\label{rbj}%
\end{align}
At this point, it is beneficial to recall that, for $n\times n$ complex
matrices -- the operators of the Hilbert space ${\mathbb{C}}^{n}$ --  the
adjoint is the transpose of the complex conjugate,
\begin{equation}
M^{\dag}=\left( M^{*}\right) ^{{\mathsf{T}}}.\label{rbk}%
\end{equation}
Noting also that, with the aid of the standard complex conjugation operator
$J$, which we have already met in the context of complex functions
[cf.~(\ref{rbi})], we can also express $M^{*}$ as
\begin{equation}
M^{*}=JMJ^{-1},\label{rbl}%
\end{equation}
(\ref{rbk}) can be reformulated as
\begin{align}
M^{\dag} & =\left( JMJ^{-1}\right) ^{{\mathsf{T}}},\label{rbm}%
\end{align}
yielding
\begin{align}
M^{{\mathsf{T}}} & =\left( JMJ^{-1}\right) ^{\dag}.\label{rbn}%
\end{align}
This enables us to define the transpose of an operator of an arbitrary Hilbert
space, with respect to a fixed conjugation operator $C$, as
\begin{align}
V^{{\mathsf{T}}} & :=\left( CVC^{-1}\right) ^{\dag}.\label{rby}%
\end{align}
This notation makes it possible to re-express the condition (\ref{rbj}) as
\begin{equation}
V=UV^{{\mathsf{T}}}U^{-1}.\label{rbo}%
\end{equation}
Why this form is worth considering becomes apparent when we turn towards the
mathematical literature. Indeed, there is a recently increasing interest in
conjugation operators and, more generally, in antiunitary ones
\cite{neretin01,Garcia2006,Chevrot2007,Tener2008,GarTen09,Garcia2009,Zagorodnyuk2010}%
. Especially relevant to our situation is the paper by \textcite{GarTen09},
where the authors study exactly condition (\ref{rbo}), in Hilbert spaces
${\mathbb{C}}^{n}$, \textit{i.e.,} for $n\times n$ complex matrices $V$, $U$.
They present a necessary and sufficient condition for those $V$'s which
fulfill
(\ref{rbo}) with some appropriate unitary $U$ -- in other words, which $V$'s
are unitarily equivalent to their transpose. Here, let us only mention some
simple special cases. To this end, we recall that a complex matrix $S$ is
a self-transpose one (also called complex symmetric) if
\begin{equation}
S^{{\mathsf{T}}}=S.\label{rbp}%
\end{equation}
It is evident that self-transpose matrices are unitarily equivalent to their
transpose. What turns out is that any matrix $Z$ that is unitarily equivalent
to a self-transpose matrix,
\begin{equation}
\left( \check{U}Z\check{U}^{-1}\right) ^{{\mathsf{T}}}= \check{U}Z\check
{U}^{-1} \quad \mbox {\big (for some unitary $\check {U}$\big )},\label{rbw}%
\end{equation}
also proves to satisfy $Z=UZ^{{\mathsf{T}}}U^{-1}$, with $U:=\big (\check
{U}^{{\mathsf{T}}}\check{U}\big )^{-1}$.

Similarly can one find that antiskew-self-transpose (also called complex
antiskewsymmetric) matrices, \textit{i.e.,} matrices of the block matrix form
\begin{equation}
A=\left(
\begin{matrix}
B_{1} & B_{2}\\
B_{3} & B_{4}%
\end{matrix}
\right) , \quad B_{2}^{{\mathsf{T}}}=-B_{2}^{}, \quad B_{3}^{{\mathsf{T}}%
}=-B_{3}^{}, \quad B_{4}^{{\mathsf{T}}}=B_{1}^{},\label{rbq}%
\end{equation}
and, more generally, matrices that are unitarily equivalent to an
antiskew-self-transpose one, are also unitarily equivalent to their transpose.

Now, the necessary and sufficient condition of the finite dimensional case
found in the work by \textcite{GarTen09} can be directly applicable for our
infinite dimensional Hilbert space situation when, for practical purposes and
applications, we restrict our search for $U$ (our search for $K$) to some
special product form. Actually, it is just such a type of restriction that we
are going to consider in the following sections. In addition, the known finite
dimensional result paves the way for the corresponding infinite dimensional
theorem to come. In fact, the straightforward infinite dimensional extension
of the finite dimensional condition -- replacing the building block matrices
involved by infinite dimensional operators -- provides a sufficient
requirement. It is plausible to expect, but is yet to be justified, that this
extended condition is not only sufficient but necessary as well.

To summarize, these mathematical results help physics to find systems with a
reciprocity property.

As an example, we close this section on the mathematical literature by quoting
a finding by \textcite{GarTen09}: For $n \le7$, any $n\times n$ complex matrix
that is unitarily equivalent to its transpose, $V=UV^{{\mathsf{T}}}U^{-1}$,
can be brought into self-transpose form by a certain unitary transformation,
$\check{U}V\check{U}^{-1}=\left( \check{U}V\check{U}^{-1}\right)
^{{\mathsf{T}}}$. In the light of Part~\ref{rab}, it will be apparent that
many physical applications benefit from this, since the labor of finding a
reciprocity operator is reduced to finding an orthonormal basis transformation
that makes the potential matrix/operator self-transpose.

\subsection{Reciprocity violation\label{rbv}}

In situations where (\ref{free-com}) and (\ref{rah}) are valid but
(\ref{CdefV}) does not hold, it is an interesting question that ``to what
extent'' reciprocity is violated. For generic $K$, this is not easy to answer
but, for the cases when $K^{2}$ commutes with $V$,
\begin{equation}
K^{2}VK^{-2}=V,\label{rce}%
\end{equation}
it is possible to provide a quantitative solution. For example, $K$'s obeying
\begin{equation}
K^{2}=e^{i\kappa}I\label{rcd}%
\end{equation}
belong to this class, and (\ref{rcd}) includes the antiunitary operators that
occur most frequently in physics, \textit{i.e.,} time reversal and charge
conjugation. [Actually, this $e^{i\kappa}$ can only be $\pm1$, as one finds
substituting it into $K K^{2}=K^{2}K$ \cite{neretin01}].

The advantage to have (\ref{rce}) is that it is equivalent to
\begin{equation}
KVK^{-1}=K^{-1}VK.\label{rcf}%
\end{equation}
Therefore, if we decompose $V$ as
\begin{equation}
V=V_{+}+V_{-},\qquad V_{\pm}:=\frac{1}{2}\left( V\pm K^{-1}V^{\dag}K\right)
\label{rcg}%
\end{equation}
then (\ref{rcf}) ensures that
\begin{equation}
KV_{\pm}K^{-1}=\pm V_{\pm}^{\dag},\label{rci}%
\end{equation}
as is easy to check. Taking a look at (\ref{CdefV}), we are pleased to realize
that we have succeeded in decomposing $V$ as a sum of a reciprocity preserving
term ($V_{+}$) and a maximally reciprocity violating one ($V_{-} $). Utilizing
(\ref{twopot}) for the $\alpha \rightarrow \beta$ and $\overline{\beta
}\rightarrow$ $\overline{\alpha}$ amplitudes yields
\begin{align}
\big \langle \beta \big \vert T\big \vert \alpha \big \rangle  &
=\big \langle \beta \big \vert T_{+}\big \vert \alpha \big \rangle +\left(
\chi_{+,\beta}^{T-},V_{-}\chi_{\alpha}^{+}\right) ,\\
\left \langle \overline{\alpha}\big \vert T\big \vert \overline{\beta
}\right \rangle  & =\left \langle \overline{\alpha}\big \vert T_{+}%
\big \vert \overline{\beta}\right \rangle +\left( \chi_{+,\overline{\alpha}%
}^{T-},V_{-}\chi_{\overline{\beta}}^{+}\right) .
\end{align}
The component $V_{+}$ is reciprocity preserving so
\begin{equation}
\big \langle \beta \big \vert T_{+}\big \vert \alpha
\big \rangle =\big \langle \overline{\alpha}\big \vert T_{+}\left \vert
\overline{\beta}\right \rangle ,\label{rck}%
\end{equation}
and thus we arrive at
\begin{equation}
\big \langle \beta \big \vert T\big \vert \alpha \big \rangle -\left \langle
\overline{\alpha}\big \vert T\big \vert \overline{\beta}\right \rangle =\left(
\chi_{+,\beta}^{T-},V_{-}\chi_{\alpha}^{+}\right) -\left( \chi_{+,\overline
{\alpha}}^{T-},V_{-}\chi_{\overline{\beta}}^{+}\right) .\label{reciviola}%
\end{equation}
Here, it is the right hand side that measures the extent to which reciprocity
is violated.

In case we wish to evaluate (\ref{reciviola}) approximately and apply the Born
approximation based formula (\ref{rdp}) then reciprocity violation is
approximated as simply as
\begin{equation}
\big \langle \beta \big \vert T\big \vert \alpha \big \rangle -\left \langle
\overline{\alpha}\left \vert T\right \vert \overline{\beta}\right \rangle
\approx \big (u_{\beta},V_{-}u_{\alpha}\big )-\big (u_{\overline{\alpha}}%
,V_{-}u_{\overline{\beta}}\big ).\label{rcx}%
\end{equation}
After some straightforward algebra on the second term of the rhs that is based
upon the observation that
\begin{equation}
K^{-1}V_{\pm}K=\pm V_{\pm}^{\dag}\label{rfe}%
\end{equation}
is also implied by (\ref{rcf}), (\ref{rcx}) can be rewritten as
\begin{equation}
\big \langle \beta \big \vert T\big \vert \alpha \big \rangle -\left \langle
\overline{\alpha}\left \vert T\right \vert \overline{\beta}\right \rangle
\approx2 \big (u_{\beta},V_{-}u_{\alpha}\big ).\label{rey}%
\end{equation}
The effect of the reciprocity violating potential component $V_{-}$ is manifest.

\subsection{Magnitude reciprocity\label{rex}}

In many experiments, one measures magnitudes $|\left \langle \beta \left \vert
T\right \vert \alpha \right \rangle |$ rather than the transition amplitudes
$\left \langle \beta \left \vert T\right \vert \alpha \right \rangle $ themselves.
Hence, it may occur that reciprocity is violated, $\big \langle \beta
\big \vert T\big \vert \alpha \big \rangle
\not =\left \langle \overline{\alpha}\left \vert T\right \vert \overline{\beta
}\right \rangle $, but this is not observed because
\begin{align}
\big \vert \big \langle \beta \big \vert T\big \vert \alpha
\big \rangle \big \vert  & =\left \vert \left \langle \overline{\alpha}
\left \vert T\right \vert \overline{\beta}\right \rangle \right \vert \label{rff}%
\end{align}
holds. Situations (\ref{rff}) can be termed quasireciprocity or
\textit{magnitude reciprocity}. Similarly, two systems may be magnitude
reciprocal partners of each other [a generalization of Sect.~\ref{rer}].

A useful connection between (full, true, complete) reciprocity and magnitude
reciprocity can be established as follows. Let us perform a unitary
transformation $\hat{U}$ on a system with potential $V$, by such a $\hat{U}$
that commutes with $H_{0}$. Completely analogously to the lines (\ref{raj}%
)--(\ref{rej}) and (\ref{rai}) can one obtain
\begin{equation}
\hat{U}G^{\pm}_{{\hskip-.02em\scriptscriptstyle E}}\hat{U}^{-1}={\hat{G}}%
^{\pm}_{{\hskip-.02em\scriptscriptstyle E}}, \quad \; \; \hat{U}\chi_{\alpha
}^{\pm}=\hat{\chi}_{\hat{\alpha}}^{\pm}, \quad \; \; \hat{U}\chi_{\beta}^{\pm
}=\hat{\chi}_{\hat{\beta}}^{\pm},\label{rel}%
\end{equation}
where ${\hat{G}}^{\pm}_{{\hskip-.02em\scriptscriptstyle E}}$ is the Green's
operator for the system with
\begin{equation}
\hat{V}:=\hat{U}V{\hat{U}}^{-1}\label{ren}%
\end{equation}
and $\hat{ \chi}_{\hat{\alpha}}^{\pm}$ etc. belong to $u_{\hat{\alpha}}%
:=\hat{U}u_{\alpha}$ etc., in system $\hat{V}$. Now let us assume that
$\hat{U}$ admits $u_{\alpha}$, $u_{\beta}$ as eigenvectors:
\begin{equation}
u_{\hat{\alpha}}\equiv \hat{U} u_{\alpha} =e^{i {\hat{\delta}}_{\alpha}%
}u_{\alpha}, \qquad u_{\hat{\beta}}\equiv \hat{U} u_{\beta} =e^{i {\hat{\delta
}}_{\beta}}u_{\beta}.\label{reo}%
\end{equation}
Then we find
\begin{equation}
\hat{\chi}_{\hat{\alpha}}^{\pm}=e^{i {\hat{\delta}}_{\alpha}} \big (I+{\hat
{G}}^{\pm}_{{\hskip-.02em\scriptscriptstyle E}}\hat{V}\big )u_{\alpha}= e^{i
{\hat{\delta}}_{\alpha}}\hat{\chi}_{\alpha}^{\pm},\label{rep}%
\end{equation}
and, proceeding similarly to (\ref{rbg}), obtain
\begin{align}
\left \langle \beta \left \vert T\right \vert \alpha \right \rangle  & =\left(
u_{\beta}^{},V\chi_{\alpha}^{+}\right)  =\left( \hat{U}u_{\beta}^{},\hat
{U}V\chi_{\alpha}^{+}\right) \nonumber \\
& =\left( u_{\hat{\beta}},\hat{V}\hat{\chi}_{\hat{\alpha}}^{+}\right)  =e^{-i
{\hat{\delta}}_{\beta}} e^{i {\hat{\delta}}_{\alpha}} \left( u_{\beta}^{}%
,\hat{V}\hat{\chi}_{\alpha}^{+}\right) \nonumber \\
& =e^{i \left( {\hat{\delta}}_{\alpha}-{\hat{\delta}}_{\beta}\right) }
\big \langle \beta \big \vert \hat{T}\big \vert \alpha \big \rangle .\label{req}%
\end{align}
We can see that the transition amplitude for the same process in the two
systems is the same in magnitude.

Therefore, if not $V$ but such a $\hat{V}$ is in a reciprocal partnership with
a $\overline{V}$ then $V$ is in magnitude reciprocal partnership with
$\overline{V}$. And, reversely, a magnitude reciprocity can be converted to
reciprocity with an appropriate transformed system.

This observation will find an important application in Sect.~\ref{rfa}, the
importance being illustrated in Part~\ref{rac}.

\section{Reciprocity for two spin/polarization degrees of freedom\label{rab}}

For waves described by a scalar square integrable complex function, and for
multicomponent wave functions with a Hamiltonian that is independent of the
spin/po\-lar\-i\-za\-tion degree of freedom, the complex conjugation $J$
usually satisfies the reciprocity conditions (\ref{free-com}) and
(\ref{CdefV}). The simplest case where the existence of a reciprocity theorem
is nontrivial is when we have a two-component wave function and the
Hamiltonian does depend on the two-component degree of freedom (in addition to
space dependence).

\subsection{Two-component wave functions and the reciprocity
conditions\label{rcy}}

In the rest of the paper, we concentrate on two-component wave functions
\begin{equation}
\psi=\left(
\begin{matrix}
\psi_{1}\\
\psi_{2}%
\end{matrix}
\right) ,\label{ran}%
\end{equation}
where $\psi_{1},\psi_{2}$ are spatially square integrable complex functions.
Such wave functions appear in the quantum mechanics of a $1/2$-spin particle,
and the two polarizations of light can also be described this way (see more on
this in Part~\ref{rac}). Let the scattering potential $V=V({\mathbf{r}})$ be a
$2\times2$ matrix valued function so the matrix entries $V_{jk}\left(
{\mathbf{r}}\right) $ ($j,k=1,2$) are complex valued functions. On the other
side, let us assume that $H_{0}$ does not mix the two components of the wave
function. For simplicity, we take the specific choice
\begin{equation}
H_{0}:=-\Delta;\label{rcm}%
\end{equation}
in case we describe a $1/2$-spin quantum particle of mass $\mu$, energy values
are considered hereafter rescaled by $2\mu/\hbar^{2}$ [$\hbar^{2}/(2\mu
)\equiv1$ convention].

A complete set of eigenfunctions of $H_{0}$ is formed by the functions
\begin{equation}
u_{\alpha}({\mathbf{r}})=p_{\alpha}e^{i{\mathbf{k}}_{\alpha} {\mathbf{r}}%
}\label{rda}%
\end{equation}
with eigenvalue
\begin{equation}
E_{\alpha}^{}={k}_{\alpha}^{2},\label{reg}%
\end{equation}
where ${\mathbf{k}}_{\alpha}$ is an arbitrary wave vector and $p_{\alpha}$ is
a vector with two complex components $p_{\alpha,j}$ ($j=1,2$), two linearly
independent such polarization vectors $p$ being considered for any fixed wave
vector. Therefore, in this concrete setting, the index $\alpha$ comprises the
eigenfunction identifying quantities as
\begin{equation}
\alpha \equiv{\mathbf{k}}_{\alpha},p_{\alpha}.\label{rds}%
\end{equation}

For describing scattering processes, it is often beneficial to introduce the
retarded ($+$) and advanced ($-$) scattering amplitudes of the original and
the adjoint problem, respectively \cite{SCHIFF}:
\begin{align}
f^{\pm}\left( {\mathbf{k}}_{\beta},\alpha \right) _{j}  & :=-\frac{1}{4\pi}\int
e^{\mp i{\mathbf{k}}_{\beta}{\mathbf{r}}} V_{jk}\left( {\mathbf{r}}\right)
\chi_{\alpha,k}^{\pm}\left(  {\mathbf{r}}\right) {\mathrm{d}}^{3}{\mathbf{r}%
},\nonumber \\
f^{T\pm}\left( {\mathbf{k}}_{\beta},\alpha \right) _{j}  & :=-\frac{1}{4\pi
}\int e^{\mp i{\mathbf{k}}_{\beta}{\mathbf{r}}}V_{jk} ^{\dagger}\left(
{\mathbf{r}}\right) \chi_{\alpha,k}^{T\pm}\left(  {\mathbf{r}}\right)
{\mathrm{d}}^{3}{\mathbf{r}},\label{scatamplT+-}%
\end{align}
each scattering amplitude being a two-component quantity ($j=1,2$), and
summation over repeated indices understood. With these notations,
(\ref{trans_ampl}) can be rewritten as
\begin{align}
\left \langle \beta \left \vert T\right \vert \alpha \right \rangle  &  =
-4\pi \left( p_{\beta},f^{+}\left( {\mathbf{k}}_{\beta}, \alpha \right) \right)
\nonumber \\
&  = -4\pi \left( f^{T-}\left( -{\mathbf{k}}_{\alpha},\beta \right) , p_{\alpha
}\right) ,\label{frametransampl}%
\end{align}
the notation $\,(\,,\,)\,$ used for the scalar product in ${\mathbb{C}}^{2}$
as well, \textit{i.e.,} $(p,q)=p_{j}^{*}q_{j}^{}=p_{1}^{*}q_{1}^{}+p_{2}%
^{*}q_{2}^{}$.

Consequently, when the reciprocity theorem (\ref{rbg}) holds then it can be
expressed via the scattering amplitudes as
\begin{align}
& \left( p_{\beta},f^{+}\left( {\mathbf{k}}_{\beta},\alpha \right)  \right)
=\left( p_{\overline{\alpha}},f^{+}\left(  {\mathbf{k}}_{\overline{\alpha}%
},\overline{\beta}\right) \right) \nonumber \\
& \quad=\left( f^{T-}\left( -{\mathbf{k}}_{\alpha},\beta \right) , p_{\alpha
}\right)  =\big (f^{T-}\big (-{\mathbf{k}}_{\overline{\beta}},\overline
{\alpha}\big ),p_{\overline{\beta}}\, \big ).\label{rei}%
\end{align}
the latter line following from (\ref{frametransampl}).

As has been mentioned in Sect.~\ref{rat}, the antiunitary operators are very
many in number. In what follows, we will restrict our attention to an
important class of antiunitary operators, namely, to those $K$ which are of
the form
\begin{equation}
K=UJ,\label{rap}%
\end{equation}
where $U$ is a $2\times2$ unitary matrix -- mixing the two components of our
two-component wave functions, in a space-independent way -- and $J$ is the
antiunitary operator of complex conjugation [cf.~(\ref{rbi})]. So to say,
these are essentially $2\times2$ antiunitary operators. With such a $K$,
(\ref{rah}) gets concretized as
\begin{align}
u_{\overline{\alpha}}({\mathbf{r}})=(Ku_{\alpha})({\mathbf{r}}) = \left(
Up_{\alpha}^{*}\right) e^{-i{\mathbf{k}}_{\alpha}{\mathbf{r}}},\nonumber \\
\rule{0em}{3ex} {\mathbf{k}}_{\overline{\alpha}}=-{\mathbf{k}}_{\alpha},
\qquad p_{\overline{\alpha}}^{} = Up_{\alpha}^{*}, \hskip4ex\label{rdr}%
\end{align}
which can also be written as
\begin{equation}
\overline{\alpha}=-{\mathbf{k}}_{\alpha},Up_{\alpha}^{*}\label{rdc}%
\end{equation}
[cf.~(\ref{rds})].

Clearly, a $K$ of this specialized form commutes with $H_{0}=-\Delta$ so the
first of the reciprocity requirements, (\ref{free-com}), is fulfilled. In
parallel, the other -- and much more nontrivial -- reciprocity condition,
(\ref{CdefV}), simplifies to
\begin{align}
V & =UV^{{\mathsf{T}}}U^{-1},\label{raq}%
\end{align}
as we have already seen at (\ref{rbo}). Note that, from the point of view of
Sect.~\ref{rdd}, now $C=J$ and $V^{{\mathsf{T}}}$ is directly the standard
matrix transpose of $V$.

Before starting to analyze condition (\ref{raq}), we make two remarks in
passing. The first is that, should $V$ be a nonlocal potential, \textit{i.e.,}
an operator acting on wave functions as
\begin{equation}
(V\psi)({\mathbf{r}})=\int V\left( {\mathbf{r}},{\mathbf{r}}^{\prime}\right)
\psi({\mathbf{r}}^{\prime}){\mathrm{d}}^{3}{{\mathbf{r}}^{\prime}},\label{rfi}%
\end{equation}
requirement (\ref{raq}) is concretized as
\begin{align}
V\left( {\mathbf{r}}, {\mathbf{r}}^{\prime}\right)  & =UV\left( {\mathbf{r}%
}^{\prime}, {\mathbf{r}}\right) U^{-1},\label{rfj}%
\end{align}
since
\begin{align}
V^{\dag}\left( {\mathbf{r}}, {\mathbf{r}}^{\prime}\right)  & = V\left(
{\mathbf{r}}^{\prime},{\mathbf{r}}\right) ^{*}.\label{rfk}%
\end{align}
Reciprocity for nonlocal potentials has been considered in
\cite{Xie2009,Xie2009b,Xie2008}.

The second remark is that a larger family of possible reciprocity operators is
also allowed: Those when $U$ of (\ref{rap}) is a Hilbert space operator acting
on wave functions as
\begin{align}
(U\psi)({\mathbf{r}}) & =Q\psi \left( {\mathbf{O}}^{-1}{\mathbf{r}} \right)
\label{rfg}%
\end{align}
with a unitary $2\times2$ matrix $Q$ and an orthogonal transformation
${\mathbf{O}}$. Such more general $K$s also commute with $H_{0}=-\Delta$, and
allow for a broader range of applications. The thorough discussion of such
operators is, however, more complicated so here we focus on the case of purely
$2\times2$ matrix $U$s.

\subsection{Unitary equivalence of a two-by-two matrix to its
transpose\label{rcz}}

Condition (\ref{raq}) says that
\begin{align}
V({\mathbf{r}}) & =UV^{{\mathsf{T}}}({\mathbf{r}})U^{-1}\label{rfh}%
\end{align}
has to be fulfilled at any location ${\mathbf{r}}$, with a certain $2\times2$
unitary matrix $U$. As a first step in analyzing this requirement, let us
consider it at a fixed ${\mathbf{r}}$. In other words, let us first pretend
that our potential is a constant.

Then the question is that which $2\times2$ matrices $V$ are unitarily
equivalent to their transpose $V^{{\mathsf{T}}}$. Let us now answer this question.

A useful characterization of any $2\times2$ matrix $V$ is done by four complex
numbers $v_{0},v_{1},v_{2},v_{3}$ via the following expansion:
\begin{align}
V  &  = v_{0}\sigma_{0}+v_{1}\sigma_{1}+v_{2}\sigma_{2}+v_{3}\sigma
_{3}\nonumber \\
&  = v_{0}\sigma_{0}+{\mathbf{v}}{\boldsymbol{\sigma}} =\left(
\begin{array}
[c]{cc}%
v_{0}+v_{3} & v_{1}-iv_{2}\\
v_{1}+iv_{2} & v_{0}-v_{3}%
\end{array}
\right) \label{caa}%
\end{align}
with $v_{0}=\frac{1}{2}{\mathrm{tr}}V$ and ${\mathbf{v}}=\frac{1}%
{2}{\mathrm{tr}}({\boldsymbol{\sigma}} V)$, where the matrices
\begin{align}
&  & \sigma_{0}  &  = \left(
\begin{matrix}
1 & 0\\
0 & 1
\end{matrix}
\right)  , & \sigma_{1}  &  = \left(
\begin{matrix}
0 & 1\\
1 & 0
\end{matrix}
\right)  , &  & \nonumber \\
&  & \sigma_{2}  &  = \left(
\begin{matrix}
0 & -i\\
i & 0
\end{matrix}
\right)  , & \sigma_{3}  &  = \left(
\begin{matrix}
1 & 0\\
0 & -1
\end{matrix}
\right)  , &  & \label{cab}%
\end{align}
\textit{i.e.,} the $2\times2$ identity matrix and the Pauli matrices, are each
self-adjoint. Here, ${\mathbf{v}}$ is frequently referred to as the
Poincar\'{e} vector corresponding to $V$. In the case of a self-adjoint $V$,
all the coefficients $v_{0},\ldots v_{3}$ are real, and, thence, ${\mathbf{v}%
}$ is a real three-dimensional vector, a vector in ${\mathbb{R}}^{3}$. When
$V$ describes absorption, too, then the imaginary part of ${\mathbf{v}}$,
again a vector in ${\mathbb{R}}^{3}$, is nonzero. Furthermore, for a unitary
$U$, this decomposition can be shown -- \textit{e.g.,} using \cite[p546]%
{MessiahII} -- to read
\begin{equation}
U=e^{i\delta}\left( \cos \frac{\varphi}{2}\sigma_{0}-i\sin \frac{\varphi}%
{2}{\mathbf{n}}{\boldsymbol{\sigma}}\right) ,\quad0\le \delta,\varphi
<2\pi,\label{rar}%
\end{equation}
where ${\mathbf{n}}$ is a real unit three-vector (element of ${\mathbb{R}}%
^{3}$).

Now, the key observation for our purposes is that, if $X$ is a matrix with
real components $(x_{0},x_{1},x_{2},x_{3})\equiv(x_{0},{\mathbf{x}})$, then
\begin{equation}
UXU^{-1}\equiv U\left( x_{0}\sigma_{0}+{\mathbf{x}}{\boldsymbol{\sigma}%
}\right) U^{-1}=x_{0}\sigma_{0}+\left( {\mathbf{O}}_{{\mathbf{n}},\varphi
}{\mathbf{x}}\right) {\boldsymbol{\sigma}},\label{ras}%
\end{equation}
where ${\mathbf{O}}_{{\mathbf{n}},\varphi}$ is the rotation of real
three-component vectors around ${\mathbf{n}}$ by angle $\varphi$, both defined
by (\ref{rar}). Eq.~(\ref{ras}) can be proven applying \cite{MessiahII}
\begin{equation}
({\mathbf{a}}{\boldsymbol{\sigma}})({\mathbf{b}}{\boldsymbol{\sigma}%
})=({\mathbf{a}}{\mathbf{b}}) \sigma_{0}+i({\mathbf{a}}\times{\mathbf{b}%
}){\boldsymbol{\sigma}}\label{rde}%
\end{equation}
and the geometrically easy-to-check formula
\begin{equation}
{\mathbf{O}}_{{\mathbf{n}},\varphi}{\mathbf{x}}=\left( {\mathbf{n}}%
{\mathbf{x}}\right) {\mathbf{n}}+\left[ {\mathbf{x}}-\left( {\mathbf{n}%
}{\mathbf{x}}\right) {\mathbf{n}}\right] \cos \varphi+\left( {\mathbf{n}}%
\times{\mathbf{x}}\right) \sin \varphi \label{rdf}%
\end{equation}
[telling that a rotation does not change the component parallel to the axis
(first term), and rotates the component orthogonal to the axis, in the plane
orthogonal to the axis (second and third terms)].

Being ready to turn towards the question of $V=UV^{{\mathsf{T}}}U^{-1}$, let
us observe that
\begin{align}
V^{{\mathsf{T}}}  &  =\left( v_{0}\sigma_{0}+v_{1}\sigma_{1}+v_{2}\sigma
_{2}+v_{3}\sigma_{3}\right) ^{{\mathsf{T}}}\nonumber \\
&  = v_{0}\sigma_{0}+v_{1}\sigma_{1}-v_{2}\sigma_{2}+v_{3}\sigma
_{3},\label{rdg}%
\end{align}
since $\sigma_{0}$, $\sigma_{1}$, and $\sigma_{3}$ are self-transpose and
$\sigma_{2}$ is anti-self-transpose. Geometrically, transposition means a
reflection of both $\,{\mathrm{Re}}\,{\mathbf{v}}\,$ and $\,{\mathrm{Im}%
}\,{\mathbf{v}}\,$ in ${\mathbb{R}}^{3}$, with respect to the $\sigma_{1}%
$--$\sigma_{3}$ plane. In notation, $V\mapsto V^{{\mathsf{T}}}$ means
\begin{equation}
{\mathrm{Re}}\,{\mathbf{v}}\mapsto{\mathbf{P}}_{13}\left( {\mathrm{Re}%
}\,{\mathbf{v}}\right) ,\qquad{\mathrm{Im}}\,{\mathbf{v}}\mapsto{\mathbf{P}%
}_{13}\left( {\mathrm{Im}}\,{\mathbf{v}}\right) ,\label{rdh}%
\end{equation}
${\mathbf{P}}_{13}$ standing for the reflection in question.

To summarize, the problem whether a $V$ admits a $U$ with which
$V=UV^{{\mathsf{T}}}U^{-1}$ is translated to the geometric question whether a
rotation exists that brings $\,{\mathbf{P}}_{13}\left( {\mathrm{Re}%
}\,{\mathbf{v}}\right) \,$ back to $\,{\mathrm{Re}}\,{\mathbf{v}}\,$ and
$\,{\mathbf{P}}_{13}\left( {\mathrm{Im}}\,{\mathbf{v}}\right) \,$ back to
$\,{\mathrm{Im}}\,{\mathbf{v}}\,$.

For the answer, first let us observe that, as has been mentioned above, a
rotation rotates the component orthogonal to the axis, in the plane orthogonal
to the axis and preserves the component parallel to the axis, thus the
difference $\:{\mathbf{O}}_{{\mathbf{n}},\varphi}{\mathbf{x}}-{\mathbf{x}}\:$
is orthogonal to ${\mathbf{n}}$ (lies in the plane orthogonal to ${\mathbf{n}%
}$). Since the difference $\,{\mathrm{Re}}\,{\mathbf{v}}-{\mathbf{P}}%
_{13}\left( {\mathrm{Re}}\,{\mathbf{v}}\right) \,$ is perpendicular to the
$\sigma_{1}$--$\sigma_{3}$ plane, the axis of any rotation that brings
$\,{\mathbf{P}}_{13}\left( {\mathrm{Re}}\,{\mathbf{v}}\right) \,$ to
$\,{\mathrm{Re}}\,{\mathbf{v}}\,$ must be within the $\sigma_{1}$--$\sigma
_{3}$ plane.

Now, a rotation ${\mathbf{O}}_{{\mathbf{n}},\varphi}$ rotates any plane that
contains ${\mathbf{n}}$ to another plane containing ${\mathbf{n}}$, the
included angle between the two planes being $\varphi$.

Consequently, to any ${\mathbf{n}}$ within the $\sigma_{1}$--$\sigma_{3}$
plane, let us consider the plane spanned by ${\mathbf{n}}$ and $\,{\mathbf{P}%
}_{13}\left( {\mathrm{Re}}\,{\mathbf{v}}\right) \,$ (which is unique as long
as ${\mathrm{Re}}\,{\mathbf{v}}\ne{\mathbf{0}}$). Let $\vartheta$ denote the
angle between this plane and the $\sigma_{1}$--$\sigma_{3}$ plane. Then the
rotation ${\mathbf{O}}_{{\mathbf{n}},\varphi}$ with $\varphi:=2\vartheta$
brings the plane of ${\mathbf{n}}$ and $\,{\mathbf{P}}_{13}\left(
{\mathrm{Re}}\,{\mathbf{v}}\right) \,$ to its $\sigma_{1}$--$\sigma_{3}%
$-reflected, \textit{i.e.,} the plane of ${\mathbf{n}}$ and $\,{\mathrm{Re}%
}\,{\mathbf{v}}\,$, and brings no other plane to its $\sigma_{1}$--$\sigma
_{3}$-reflected.

Hence, if we want $\,{\mathbf{P}}_{13}\left( {\mathrm{Im}}\,{\mathbf{v}%
}\right) \,$ to be rotated to $\,{\mathrm{Im}}\,{\mathbf{v}}\,$, too, then
$\,{\mathbf{P}}_{13}\left( {\mathrm{Im}}\,{\mathbf{v}}\right) \,$ must be
contained in the plane of ${\mathbf{n}}$ and $\,{\mathbf{P}}_{13}\left(
{\mathrm{Re}}\,{\mathbf{v}}\right) \,$. We can always choose such an
${\mathbf{n}}$ that this is satisfied:\newline(1) If the plane spanned by
$\,{\mathbf{P}}_{13}\left( {\mathrm{Re}}\,{\mathbf{v}}\right) \,$ and
$\,{\mathbf{P}}_{13}\left( {\mathrm{Im}}\,{\mathbf{v}}\right) \,$ intersects
with the $\sigma_{1}$--$\sigma_{3}$ plane at a line then let ${\mathbf{n}}$ be
along this line; \newline(2) if ${\mathrm{Re}}\,{\mathbf{v}}$ and
${\mathrm{Im}}\,{\mathbf{v}}$ are the multiples of each other, outside the
$\sigma_{1}$--$\sigma_{3}$ plane, then any ${\mathbf{n}}$ within the
$\sigma_{1}$--$\sigma_{3}$ plane suffices;  \newline(3) at last, if both
${\mathrm{Re}}\,{\mathbf{v}}$ and ${\mathrm{Im}}\,{\mathbf{v}}$ are within the
$\sigma_{1}$--$\sigma_{3}$ plane then no rotation is needed (we choose the
identity transformation).

We have concluded that any $V$ admits a $U$ that can be chosen based on
${\mathrm{Re}}\,{\mathbf{v}}$ and ${\mathrm{Im}}\,{\mathbf{v}}$, the rotation
${\mathbf{O}}_{{\mathbf{n}},\varphi}$ corresponding to $U$ being determined
uniquely if ${\mathrm{Re}}\,{\mathbf{v}}$ and ${\mathrm{Im}}\,{\mathbf{v}}$
are linearly independent.

Having answered our problem, we can make the following useful `by-product'
observation. If we rotate by $-\vartheta$ instead of $2\vartheta$
then we rotate both ${\mathrm{Re}}\,{\mathbf{v}}$ and ${\mathrm{Im}%
}\,{\mathbf{v}}$ into the $\sigma_{1}$--$\sigma_{3}$ plane. In other words, we
establish a unitary transformation of $V$ into a self-transpose matrix.
This transformation embodies an example of $\check{U}$ in Sect.~\ref{rdd}.

The fact that any $2\times2$ matrix is unitarily equivalent to a
self-transpose one has been known for a long time -- see the mathematical
literature invoked in Sect.~\ref{rdd} -- and here it emerges in a geometric incarnation.

Furthermore, it is adequate to recall again the results mentioned in
Sect.~\ref{rdd}, revealing that, in dimensions $n \le7$, unitary equivalence
to the transpose holds for the same matrices as unitary equivalence to a
self-transpose matrix.

During the above consideration, the special case when ${\mathrm{Re}%
}\,{\mathbf{v}}$ and ${\mathrm{Im}}\,{\mathbf{v}}$ are the multiples of each
other behaved in a distinguished way. This occurs when, in the decomposition
(\ref{caa}), ${\mathbf{v}}=c{\mathbf{b}}$ with a complex $c$ and a real
three-vector ${\mathbf{b}}$:
\begin{equation}
V=v_{0}\sigma_{0}+c{\mathbf{b}}{\boldsymbol{\sigma}} .\label{rdk}%
\end{equation}
These $V$ can be called univectorial, and the other ones, with linearly
independent ${\mathrm{Re}}\,{\mathbf{v}}$ and ${\mathrm{Im}}\,{\mathbf{v}}$,
bivectorial. The family of univectorial potentials is important from the
physical point of view, as it includes the scattering interaction of a neutron
with an absorptive medium with a magnetic field -- where ${\mathbf{b}}$ is the
magnetic field ${\mathbf{B}}$ itself --  and also the M\"{o}ssbauer scattering
of photons (see Part~\ref{rac}) -- where ${\mathbf{b}}$ is not directly the
magnetic field but is determined by it. Univectorial potentials will play two
important roles in our subsequent discussion, one uncovered in Sect.~\ref{rfa}
and the other, based on their special property in the above geometric picture,
turning out in the following section.

\subsection{A criterion ensuring a reciprocity operator\label{rad}}

Now, we are prepared to make the second step, taking the space dependence of
$V$ into consideration. The corresponding second question is: When does
(\ref{rfh}) hold for any location ${\mathbf{r}}$?

Fortunately, the consideration we made in the above section enables us to give
the answer immediately. Namely, any given $U$ means a given plane -- which it
can rotate to its reflected -- so if and only if all ${\mathrm{Re}%
}\,{\mathbf{v}}({\mathbf{r}})$, ${\mathrm{Im}}\,{\mathbf{v}}({\mathbf{r}})$ of
all locations ${\mathbf{r}}$ are in a common plane within ${\mathbb{R}}^{3}$
then a $U$ we seek exists, and it is actually the one corresponding to this
common plane. Potentials with such a common plane may be called globally uniplanar.

One remarkable special case is a potential that is globally (everywhere)
univectorial [see (\ref{rdk})]. For such potentials, all those locally single
vectors ${\mathbf{b}}({\mathbf{r}})$ have to fall within a common plane.

Another example, important for applications like those considered in
Part~\ref{rac}, is when the potential is piecewise constant, taking a value
$V_{1}$ on one space domain, a value $V_{2}$ on another domain, etc. Then the
vectors
\begin{equation}
{\mathrm{Re}}\,{\mathbf{v}}_{1},\; \: {\mathrm{Im}}\,{\mathbf{v}}_{1},\; \:
{\mathrm{Re}}\,{\mathbf{v}}_{2},\; \: {\mathrm{Im}}\,{\mathbf{v}}_{2},\; \:
\ldots,\; \: {\mathrm{Re}}\,{\mathbf{v}}_{n},\; \: {\mathrm{Im}}\,{\mathbf{v}%
}_{n}\label{rdj}%
\end{equation}
must be within a common plane. Naturally, here again, rotating by half the
angle (see the previous section) means a unitary transformation of each of
these potentials to self-transpose ones.

For univectorial potential values $V_{1}$, $V_{2}$, \ldots, $V_{n}$ with
${\mathbf{b}}_{1}, {\mathbf{b}}_{2}, \ldots, {\mathbf{b}}_{n}$, this criterion
of a common plane implies that any two univectorial values admit a common
unitary transformation to their transpose (and another one to self-transpose
ones), the common plane being spanned by ${\mathbf{b}}_{1}$ and ${\mathbf{b}%
}_{2}$, being orthogonal to ${\mathbf{b}}_{1}\times{\mathbf{b}}_{2}$, with the
exception of collinear ${\mathbf{b}}_{1},{\mathbf{b}}_{2}$ when the common
plane is not even unique. The existence of a common plane becomes nontrivial
only for $n\ge3$.

We close this section by mentioning the relationship between commutativity and
joint reciprocity of $2\times2$ matrices. This question is a natural one in
the light of the well-known connection between commutativity and simultaneous
diagonalizability of diagonalizable matrices. Now, omitting straightforward
details, one can find that $V_{1}V_{2}=V_{2}V_{1}$ if and only if
\begin{align}
&  \operatorname{Re}{\mathbf{v}}_{1}\times \operatorname{Re}{\mathbf{v}}%
_{2}-\operatorname{Im}{\mathbf{v}}_{1}\times \operatorname{Im}{\mathbf{v}}%
_{2}\nonumber \\
& \quad= \operatorname{Re}{\mathbf{v}}_{1}\times \operatorname{Im}{\mathbf{v}%
}_{2}+\operatorname{Im}{\mathbf{v}}_{1}\times \operatorname{Re}{\mathbf{v}}%
_{2}={\mathbf{0}}.\label{cax}%
\end{align}
It follows that $\operatorname{Re}{\mathbf{v}}_{1}$, $\operatorname{Re}%
{\mathbf{v}}_{2}$, $\operatorname{Im}{\mathbf{v}}_{1}$ and $\operatorname{Im}%
{\mathbf{v}}_{2}$ all lie within a plane. Hence, $V_{1}$ and $V_{2}$ admit a
joint reciprocity operator. The converse direction does not hold: For example,
$V_{1}=\sigma_{1}$ and $V_{2}=\sigma_{3}$ do not commute but share a common
$U$ that connects both of them with their own transpose. In fact, $U=\left(
\begin{smallmatrix}
1 & 0\\
0 & 1
\end{smallmatrix}
\right) $ suffices as $\sigma_{1}$ and $\sigma_{3}$ are already
self-transpose. Consequently, commutativity implies, but does not follow from,
joint reciprocity.

\subsection{A special case of reciprocity: Time reversal\label{raf}}

Similarly to how $KVK^{-1}=V^{\dag}$ is simplified, via the decomposition
$K=UJ $, to $V=UV^{{\mathsf{T}}}U^{-1}$, the generalization $KVK^{-1}%
=\overline{V}^{\dag}$ seen at (\ref{ral}) becomes $\overline{V}%
=UV^{{\mathsf{T}}}U^{-1} $. A physically important application of this
reciprocal partnership between different systems, discussed in Sect.~\ref{rbx}%
, occurs for $U_{T}=-i\sigma_{2}$, when $K_{T}=U_{T}J$ is the time reversal
operator of 1/2-spin quantum mechanics \cite{SCHIFF,MessiahII}.

Using (\ref{rde}) and (\ref{rdg}), it is easy to check that the time reversal
$K_{T}$ maps
\begin{equation}
V=v_{0}\sigma_{0}+v_{1}\sigma_{1}+v_{2}\sigma_{2}+v_{3}\sigma_{3}\label{reh}%
\end{equation}
to
\begin{align}
\overline{V}=U_{T}^{\vphantom {|}}V^{{\mathsf{T}}}U_{T}^{-1} & = \sigma
_{2}\left( v_{0}\sigma_{0}+v_{1}\sigma_{1}-v_{2}\sigma_{2}+ v_{3}\sigma
_{3}\right) \sigma_{2} ^{-1}\nonumber \\
& =v_{0}\sigma_{0}-v_{1}\sigma_{1}-v_{2}\sigma_{2}- v_{3}\sigma_{3}\nonumber \\
&  = v_{0}\sigma_{0}-{\mathbf{v}}{\boldsymbol{\sigma}}.\label{rdm}%
\end{align}
Two remarkable specializations are\newline(1) when ${\mathbf{v}}={\mathbf{0}}%
$, and\newline(2) when $v_{0}=0$.

In case~(1), we find $\overline{V}=V$, which means that, for
spin/polarization-independent potentials, time reversal is a reciprocity
operator. It is essential to emphasize that $v_{0}$ is allowed be any complex
number for this so reciprocity holds true also in cases when time reversal is
not a symmetry (when ${\mathrm{Im}}\,v_{0}\ne0$). This situation serves as an
example for what Sect.~\ref{rag} has pointed out: Reciprocity is not the same
as time reversal invariance.

In case~(2), we have $\overline{V}=-V$, that is, time reversal has mapped the
potential to its negative. When $V=c{\mathbf{B}}{\boldsymbol{\sigma}}$
 -- magnetic dipole in a magnetic field -- this means $\overline{{\mathbf{B}}%
}=-{\mathbf{B}}$. This is in conformity with the well-known fact that time
reversal must be accompanied by the reversal of magnetic field to obtain a
generalized symmetry (a spectrum preserving equivalence between systems),
which is now a reciprocal partnership as well. Reciprocity is more general
than time reversal invariance.

For clarity, let us remark that the change $\overline{{\mathbf{B}}%
}=-{\mathbf{B}}$ cannot be covered by a gauge transformation (\ref{rct}).
Gauge transformations of the electromagnetic potentials always keep the
electric and magnetic fields invariant. $\overline{{\mathbf{B}}}=-{\mathbf{B}%
}$ does indeed mean two different physical systems.

\subsection{Reciprocity vs. rotation\label{Rotavsreci}}

The physical setting for testing the reciprocity theorem $\left \langle
\beta \big \vert T\big \vert \alpha \right \rangle =\left \langle \overline
{\alpha}\big \vert T\big \vert \overline{\beta}\right \rangle $ is the
following: A source emits a wave from direction ${\mathbf{k}}_{\alpha}$ with
polarization $p_{\alpha}$ towards the scattering object, and the scattered
wave is detected in direction ${\mathbf{k}}_{\beta}$, with polarization
$p_{\beta}$; and this scattering is compared to when the incoming direction is
${\mathbf{k}}_{\overline{\beta}}$ with incoming polarization $p_{\overline
{\beta}}$, and the detector direction is ${\mathbf{k}}_{\overline{\alpha}}$,
detecting polarization $p_{\overline{\alpha}}$. Now, in experiments at large
facilities it would be difficult to change the direction of the source. It is
much more feasible to perform a rotation of the small scatterer sample. Let us
now derive a version of the reciprocity theorem that does not require to
modify the incoming direction, which can be achieved with the help of an
appropriate rotation.

Let us start by repeating that, by (\ref{rdc}), we have now
\begin{equation}
{\mathbf{k}}_{\overline{\alpha}}=-{\mathbf{k}}_{\alpha},\qquad{\mathbf{k}%
}_{\overline{\beta}}=-{\mathbf{k}}_{\beta}.\label{rdq}%
\end{equation}
Recall also that we are considering elastic scattering processes (see
Sect.~\ref{rau}), which ensures $E_{\alpha}=E_{\beta}$ and implies $k_{\alpha
}=k_{\beta}$ [cf.~(\ref{reg})]. This means that the vectors ${\mathbf{k}%
}_{\alpha}$ and ${\mathbf{k}}_{\beta}$ can be interchanged by some rotation.
More closely, we are interested in a rotation that brings $\,{\mathbf{k}%
}_{\overline{\beta}}=-{\mathbf{k}}_{\beta}\,$ to ${\mathbf{k}}_{\alpha}$.
Actually, one can observe that the rotation by $\pi \equiv180^{\circ}$ around
the direction of the momentum transfer $\,{\mathbf{k}}_{\beta}-{\mathbf{k}%
}_{\alpha}\,$ realizes this desire and, at the same time, maps ${\mathbf{k}%
}_{\overline{\alpha}}=-{\mathbf{k}}_{\alpha}$ to ${\mathbf{k}}_{\beta}$, which
has the benefit that the detector direction also remains the same as for the
process $\alpha \to \beta$.

Furthermore, in the special case of forward scattering, ${\mathbf{k}}_{\beta
}={\mathbf{k}}_{\alpha}$, when there is no momentum transfer, any direction
orthogonal to ${\mathbf{k}}_{\alpha}$ suffices for the same purpose.

Let ${\mathbf{O}}_{R }$ denote this rotation around this direction and angle,
\begin{equation}
{\mathbf{n}}_{R }^{}=\left( {\mathbf{k}}_{\beta}-{\mathbf{k}}_{\alpha}\right)
/\left| {\mathbf{k}}_{\beta}-{\mathbf{k}}_{\alpha}\right| ,\qquad \varphi_{R
}^{}=\pi \label{rdt}%
\end{equation}
(except for ${\mathbf{k}}_{\beta}={\mathbf{k}}_{\alpha}$, when ${\mathbf{n}%
}_{R }$ is arbitrary up to ${\mathbf{n}}_{R }{\mathbf{k}}_{\alpha}=0$). It is
known from 1/2-spin quantum mechanics that a rotation is represented as a
Hilbert space operator on the wave functions as
\begin{equation}
\left( R \psi \right) \left( {\mathbf{r}}\right) =U_{R }^{}\psi \left(
{\mathbf{O}}_{R }^{-1}{\mathbf{r}}\right) ,\label{rdu}%
\end{equation}
where $U_{R }$ is the $SU(2)$ transformation [see also (\ref{rar})]
\begin{equation}
U_{R }=\cos \frac{\varphi_{R }}{2}\sigma_{0}-i\sin \frac{\varphi_{R }}%
{2}{\mathbf{n}}_{R }{\boldsymbol{\sigma}}=-i{\mathbf{n}}_{R }%
{\boldsymbol{\sigma}}.\label{rdw}%
\end{equation}
Especially,
\begin{equation}
\left( R u_{\overline{\alpha}}^{}\right) \left( {\mathbf{r}}\right) =U_{R }%
^{}p_{\overline{\alpha}}^{}\,e^{i\left( {\mathbf{O}}_{R }{\mathbf{k}%
}_{\overline{\alpha}}\right) \cdot{\mathbf{r}}}=U_{R }^{}Up_{\alpha}%
^{*}\,e^{-i{\mathbf{k}}_{\beta}{\mathbf{r}}}\label{rdv}%
\end{equation}
[note $\, {\mathbf{k}}\cdot \left( {\mathbf{O}}_{R }^{-1}{\mathbf{r}}\right)
=\left( {\mathbf{O}}_{R }^{}{\mathbf{k}}\right) \cdot{\mathbf{r}}\, $], giving
for the eigenfunction index [cf.~(\ref{rds}) and (\ref{rdc})]
\begin{equation}
R \overline{\alpha} \, = \, {\mathbf{O}}_{R }^{}{\mathbf{k}}_{\overline{\alpha}}
^{}\!\:,\!\,U_{R }^{}p_{\overline{\alpha}}^{} \, = \, -{\mathbf{k}}_{\beta},U_{R }%
^{}Up_{\alpha}^{*}.\label{rdx}%
\end{equation}

Our $H_{0}$ is rotation invariant so, with $V_{R }\equiv R VR ^{-1}$,
\begin{align}
R HR ^{-1} & =H_{0}+R VR ^{-1}=H_{0}+V_{R },\label{rdy}\\
R G^{\pm}_{{\hskip-.02em\scriptscriptstyle E} }R ^{-1} & =\left( {G^{\pm
}_{{\hskip-.02em\scriptscriptstyle E} }}\right) _{R },\label{rdyb}%
\end{align}
where $\left( {G^{\pm}_{{\hskip-.02em\scriptscriptstyle E} }}\right) _{R }$
denotes the Green's operator corresponding to $V_{R }$ -- see (\ref{rcw}) with
Hamiltonian $H_{0}+V_{R }$. This enables us to derive, from (\ref{Lip1}),
\begin{align}
R \chi_{\overline{\alpha}}^{\pm} & =R u_{\overline{\alpha}}^{\phantom {|}}+R
G^{\pm}_{{\hskip-.02em\scriptscriptstyle E} }R ^{-1}\left( R VR ^{-1}\right) R
u_{\overline{\alpha}} ^{\phantom{|}}\nonumber \\
& =u_{R \overline{\alpha}}^{\phantom {|}}+\left( {G^{\pm}%
_{{\hskip-.02em\scriptscriptstyle E} }}\right) _{R }^{\phantom {|}}u_{R
\overline{\alpha}}^{\phantom {|}}.\label{rdz}%
\end{align}
Consequently, the reciprocity theorem can be re-expressed first as
\begin{align}
\left \langle \beta \big \vert T\big \vert \alpha \right \rangle  & =\left \langle
\overline{\alpha}\big \vert T\big \vert \overline{\beta} \right \rangle =\!
\left( u_{\overline{\alpha}},V\chi_{\overline{\beta}}^{+}\right) \! =\! \left(
R u_{\overline{\alpha}},R V R ^{-1} R \chi_{\overline{\beta}}^{+}\right)
\!,\label{rea}%
\end{align}
and then in final form,
\begin{align}
\hskip -.7em \big \langle {\mathbf{k}}_{\beta},p_{\beta}%
\big \vert T\big \vert {\mathbf{k}}_{\alpha},p_{\alpha}\big \rangle  &
=\big \langle {\mathbf{k}}_{\beta},U_{R }Up_{\alpha}^{*}\big \vert T_{R
}\big \vert {\mathbf{k}}_{\alpha},U_{R }Up_{\beta}^{*}%
\big \rangle ,\label{reb}%
\end{align}
where the eigenfunction indices have been explicitly displayed, and $T_{R }$
stands for the transition amplitude for scattering on $V_{R }$.

One can observe that the rotation is able to transform the incoming momentum
to the outgoing one and vice versa, but unable to transform an incoming
polarization to an outgoing one (and vice versa). Reciprocity is not the same
as a rotational invariance.

It can be practical to choose the $z$ coordinate axis of our coordinate system
parallel to ${\mathbf{k}}_{\beta}-{\mathbf{k}}_{\alpha}$. Then ${\mathbf{n}%
}_{R }=(0\; 0\; 1)^{{\mathsf{T}}}$, $U_{R }=-i\sigma_{3}$. Since the
transition amplitude, being a Hilbert space scalar product, remains invariant
if we multiply both the initial and the final state by $i$, (\ref{reb}) is
simplified to
\begin{align}
\big \langle {\mathbf{k}}_{\beta},p_{\beta}\big \vert T\big \vert {\mathbf{k}%
}_{\alpha},p_{\alpha}\big \rangle  & =\big \langle {\mathbf{k}}_{\beta}%
,\sigma_{3}Up_{\alpha}^{*}\big \vert T_{R }\big \vert {\mathbf{k}}_{\alpha
},\sigma_{3}Up_{\beta}^{*}\big \rangle .\label{red}%
\end{align}
In the special case $U=\left(
\begin{smallmatrix}
1 & 0\\
0 & 1
\end{smallmatrix}
\right) $, displaying explicitly the two components of the polarization
vectors,
\begin{align}
& \left \langle {\mathbf{k}}_{\beta},{p_{\beta}}_{1},{p_{\beta}}_{2}\left \vert
T\right \vert {\mathbf{k}}_{\alpha},{p_{\alpha}}_{1},{p_{\alpha}}%
_{2}\right \rangle \nonumber \\
& \quad=\left \langle {\mathbf{k}}_{\beta},{p_{\alpha}}_{1}^{*},-{p_{\alpha}%
}_{2}^{*}\left \vert T_{R }\right \vert {\mathbf{k}}_{\alpha},{p_{\beta}}%
_{1}^{*},-{p_{\beta}}_{2}^{*}\right \rangle .\label{reci22exp}%
\end{align}

On the other side, in forward scattering, ${\mathbf{k}}_{\beta}={\mathbf{k}%
}_{\alpha}$, a distinguished choice for the $z$ axis is the direction of
${\mathbf{k}}_{\alpha}$. If ${\mathbf{n}}_{R }$ is in the $x$ direction then
$U_{R }=-i\sigma_{1}$, and the analogous result is
\begin{align}
\big \langle {\mathbf{k}}_{\beta},p_{\beta}\big \vert T\big \vert {\mathbf{k}%
}_{\alpha},p_{\alpha}\big \rangle  & =\big \langle {\mathbf{k}}_{\beta}%
,\sigma_{1}Up_{\alpha}^{*}\big \vert T_{R }\big \vert {\mathbf{k}}_{\alpha
},\sigma_{1}Up_{\beta}^{*} \big \rangle ,\label{reu}%
\end{align}
which, for $U=\left(
\begin{smallmatrix}
1 & 0\\
0 & 1
\end{smallmatrix}
\right) $, simplifies to
\begin{align}
& \left \langle {\mathbf{k}}_{\beta},{p_{\beta}}_{1},{p_{\beta}}_{2} \left \vert
T\right \vert {\mathbf{k}}_{\alpha},{p_{\alpha}}_{1}, {p_{\alpha}}%
_{2}\right \rangle \nonumber \\
& \quad=\left \langle {\mathbf{k}}_{\beta},{p_{\alpha}}_{2}^{*}, {p_{\alpha}%
}_{1}^{*}\left \vert T_{R }\right \vert {\mathbf{k}}_{\alpha},{p_{\beta}}%
_{2}^{*},{p_{\beta}}_{1}^{*}\right \rangle .\label{rev}%
\end{align}
Similarly, if ${\mathbf{n}}_{R }$ is in the $y$ direction then $U_{R
}=-i\sigma_{2}$, leading to
\begin{align}
\big \langle {\mathbf{k}}_{\beta},p_{\beta}\big \vert T\big \vert {\mathbf{k}%
}_{\alpha},p_{\alpha}\big \rangle  & =\big \langle {\mathbf{k}}_{\beta
},-i\sigma_{2}Up_{\alpha}^{*}\big \vert T_{R }\big \vert {\mathbf{k}}_{\alpha
},-i\sigma_{2}Up_{\beta}^{*} \big \rangle ,\label{rfc}%
\end{align}
and, for $U=\left(
\begin{smallmatrix}
1 & 0\\
0 & 1
\end{smallmatrix}
\right) $, to
\begin{align}
& \left \langle {\mathbf{k}}_{\beta},{p_{\beta}}_{1},{p_{\beta}}_{2}\left \vert
T\right \vert {\mathbf{k}}_{\alpha},{p_{\alpha}}_{1},{p_{\alpha}}%
_{2}\right \rangle \nonumber \\
& \quad=\left \langle {\mathbf{k}}_{\beta},-{p_{\alpha}}_{2}^{*}, {p_{\alpha}%
}_{1}^{*}\left \vert T_{R }\right \vert {\mathbf{k}}_{\alpha},-{p_{\beta}}%
_{2}^{*},{p_{\beta}}_{1}^{*}\right \rangle .\label{rfd}%
\end{align}

For reciprocal partners $V$ and $\overline{V}=UV^{{\mathsf{T}}}U^{-1} $, the
generalization of (\ref{reb}) reads, naturally,
\begin{align}
\hskip -.7em \big \langle {\mathbf{k}}_{\beta},p_{\beta}%
\big \vert T\big \vert {\mathbf{k}}_{\alpha},p_{\alpha}\big \rangle  &
=\big \langle {\mathbf{k}}_{\beta},U_{R }Up_{\alpha}^{*}\big \vert {\overline
{T}}_{R }\big \vert {\mathbf{k}}_{\alpha},U_{R } Up_{\beta}^{*}%
\big \rangle ,\label{rgb}%
\end{align}
where ${\overline{T}}$ corresponds to $\overline{V}$ and ${\overline{T}}_{R }$
to the rotated $R {\overline{V}} R ^{-1}$.

If the scattering potential $V$ is spin/polarization independent,
$V({\mathbf{r}}) = v_{0}({\mathbf{r}}) \sigma_{0}$, then not only the free
solutions $u_{\alpha}({\mathbf{r}})$ are of the form (\ref{rda}) but the
corresponding scattering solutions $\chi^{\pm}_{\alpha}({\mathbf{r}})$,
$\chi^{T\pm}_{\alpha}({\mathbf{r}})$ are also a product of the polarization
term $p_{\alpha}$ and a space dependent function, the latter being the
appropriate scattering solution of the reduced problem of scalar wave
scattering on potential $v_{0}({\mathbf{r}})$. Accordingly, the transition
amplitude is also factorizable, as
\begin{align}
\big \langle {\mathbf{k}}_{\beta},p_{\beta}\big \vert T\big \vert {\mathbf{k}%
}_{\alpha},p_{\alpha}\big \rangle  & = (p_{\beta}, p_{\alpha}) \,
\big \langle {\mathbf{k}}_{\beta}\big \vert T^{\text{red}}%
\big \vert {\mathbf{k}}_{\alpha}\big \rangle .\label{rgi}%
\end{align}
As a consequence, the reciprocity theorem combined with rotation,
\textit{viz.}, (\ref{reb}) is easily seen to get simplified to
\begin{align}
(p_{\beta}, p_{\alpha}) \, \big \langle {\mathbf{k}}_{\beta}%
\big \vert T^{\text{red}}\big \vert {\mathbf{k}}_{\alpha}\big \rangle  & =
(p_{\beta}, p_{\alpha}) \, \big \langle {\mathbf{k}}_{\beta}\big \vert T_{R}%
^{\text{red}}\big \vert {\mathbf{k}}_{\alpha}\big \rangle \label{rgj}%
\end{align}
[as $(U_{R }Up_{\alpha}^{*}, U_{R }Up_{\beta}^{*}) = (p_{\alpha}^{*},
p_{\beta}^{*}) = (p_{\beta}, p_{\alpha})$], where the rhs of (\ref{rgj})
refers to the reduced scalar scattering on the rotated scalar potential
\begin{align}
(v_{0})_{R}^{\vphantom {|}}({\mathbf{r}}) = v_{0}({\mathbf{O}}_{R }%
^{-1}{\mathbf{r}}).\label{rgk}%
\end{align}
This shows that, for polarization independent scattering -- but only in those
cases -- reciprocity acts the same way as a rotation.

Historically, reciprocity was first studied for polarization independent
phenomena. This explains why it had to be a later step to recognize that, in
polarization dependent wave scattering, reciprocity deviates from a rotation
invariance, the difference manifesting itself in the polarization degree of
freedom. Naturally, the third step, formulating reciprocity for general
quantum/wave systems, as done here in Part~\ref{raa}, makes it apparent that
reciprocity is related to an antiunitary operator and rotations to unitary
ones. This remarkable mathematical difference carries considerably different
physical content.

\subsection{Fixing the reciprocity operator to processes\label{rcv}}

Reciprocity and its violation has already been discussed at the general level,
in Part~\ref{raa}. It is interesting to observe that, for two-component wave
functions, there is a special additional possibility. Namely, if we have a
Hamiltonian $H_{0}+V$ and are interested in the relationship between any two
scattering processes $\alpha \to \beta$ and $\overline{\beta}\to \overline
{\alpha}$ then there may be a $K$, unique in the range chosen at (\ref{rap}),
which satisfies
\begin{equation}
u_{\overline{\alpha}}=Ku_{\alpha},\qquad u_{\overline{\beta}}=Ku_{\beta
}.\label{ref}%
\end{equation}
In fact, using the decomposition (\ref{rap}), the two formulae
\begin{equation}
p_{\overline{\alpha}}=Up_{\alpha}^{*},\qquad p_{\overline{\beta}}=Up_{\beta
}^{*}\label{rec}%
\end{equation}
define a linear operator $U$ uniquely -- as long as $p_{\alpha}$ and $p_{\beta
}$ are linearly independent -- and if this $U$ is unitary then we have arrived
at a uniquely defined antiunitary $K$.

Then we are allowed to ask whether $V=UV^{{\mathsf{T}}}U^{-1}$, and if not
then how much reciprocity is violated, along the lines of the general
treatment of Sect.~\ref{rbv}.

This is a reverse approach in the sense that not $u_{\alpha}$, $u_{\beta}$,
and $K$ define $u_{\overline{\alpha}}$ and $u_{\overline{\beta}}$, as we
proceeded at (\ref{rah}), but $u_{\alpha}$, $u_{\beta}$, $u_{\overline{\alpha
}}$, and $u_{\overline{\beta}}$ define $K$. This special possibility, which
exists only for two polarization degrees of freedom, enlarges the range of
application of the reciprocity theorem compared to the general case.

Actually, a very frequent approach is to fix processes -- momenta and
polarizations -- and to ask whether the corresponding $K$ is a reciprocity
operator for the system's $V$. In fact, typically (tacitly) $U=\sigma_{0}$,
$K=J$ is assumed and thus the question is the self-transposeness of $V$.

Now, for $n\le7$ polarization degrees of freedom, indeed the reciprocity
property is equivalent to self-transposeness with respect to an appropriate
orthogonal basis (cf.\ Sect.~\ref{rdd}). However, it is beneficial to be
prepared for other polarization bases. Furthermore, there are physical
situations with dimensions $n>7$, like many-particle quantum mechanics of
spin-half particles, or the presently considered two-component wave situations
when we do not restrict ourselves to $2\times2$ matrices $U$ but allow $U$ be
some essentially infinite dimensional unitary operator of the infinite
dimensional Hilbert space.

That being said, the approach to ask $2\times2$ self-transposeness fits
numerous physical settings and is a convenient and simple means of finding
situations possessing reciprocity.

As a matter of fact, an extended version is to choose two orthogonal
polarizations $p_{a}, p_{b}$ and to consider four, rather than two, processes
to relate:
\begin{equation}
p_{a}\to p_{a},\quad p_{a}\to p_{b},\quad p_{b}\to p_{a}, \quad p_{b}\to
p_{b}.\label{rfm}%
\end{equation}
(Some ${\mathbf{k}}_{\alpha}, {\mathbf{k}}_{\beta}$ must also be fixed, and a
rotation of the previous Section may also be included.) If
\begin{equation}
p_{a}=\left(
\begin{matrix}
1\\
0
\end{matrix}
\right) , \quad p_{b}=\left(
\begin{matrix}
0\\
1
\end{matrix}
\right) \label{rfn}%
\end{equation}
then the question of reciprocity is the question of whether $V_{jk}=V_{kj}$
$(j, k = 1, 2)$. Among these, the only nontrivial one is whether or not
\begin{equation}
V_{12} = V_{21}\label{rfl}%
\end{equation}
holds. This single condition ensures reciprocity for four processes.

\subsection{Potentials exhibiting magnitude reciprocity\label{rfa}}

As has been raised in Sect.~\ref{rex}, many experiments measure only the
magnitude of the transition amplitude, and one can speak of magnitude
reciprocity when two transition amplitudes, related by an antiunitary $K$, are
the same in magnitude. There, we derived a possibility for this to occur,
which consideration, actualized for the situation (\ref{rfm})--(\ref{rfl}),
implies that, if $V$ is self-transpose and
\begin{equation}
{\hat{U}}=\left(
\begin{matrix}
e^{i {\hat{\delta}}_{a}} & 0\\
0 & e^{i {\hat{\delta}}_{b}}%
\end{matrix}
\right) ,\label{rfo}%
\end{equation}
then
\begin{equation}
{\hat{V}}:={\hat{U}}V{\hat{U}}^{-1}\label{rfp}%
\end{equation}
exhibits magnitude reciprocity for $K=J$ and processes (\ref{rfm})--(\ref{rfn}).

In detail, (\ref{rfp}) reads
\begin{equation}
\left(
\begin{matrix}
{\hat{V}}_{11} & {\hat{V}}_{12}\\
{\hat{V}}_{21} & {\hat{V}}_{22}%
\end{matrix}
\right)  = \left(
\begin{matrix}
V_{11} & e^{-i \left( {\hat{\delta}}_{b} - {\hat{\delta}}_{a} \right) }
V_{12}\\
e^{i \left( {\hat{\delta}}_{b} - {\hat{\delta}}_{a}\right) } V_{21} & V_{22}%
\end{matrix}
\right) \label{rfq}%
\end{equation}
so $V_{12} = V_{21}$ implies
\begin{equation}
{\hat{V}}_{12}=e^{i {\hat{\delta}}}{\hat{V}}_{21}\label{rfr}%
\end{equation}
with some angle ${\hat{\delta}}$. This means that, similarly to that a
self-transpose potential is reciprocal with respect to $K=J$, a ``phase
self-transpose potential'' [\textit{i.e.,} a potential obeying
(\ref{rfr})] is
magnitude reciprocal with respect to $K=J$.

An example for a phase self-transpose potential is an everywhere univectorial
potential [cf.\ (\ref{rdk})], \textit{i.e.,}
\begin{equation}
V({\mathbf{r}})=v_{0}({\mathbf{r}})\sigma_{0}+ c({\mathbf{r}}){\mathbf{b}%
}({\mathbf{r}}){\boldsymbol{\sigma}}\label{rfs}%
\end{equation}
with real ${\mathbf{b}}({\mathbf{r}})$ and possibly complex $v_{0}%
({\mathbf{r}}), c({\mathbf{r}})$, if it obeys some restriction. That
restriction can be revealed via the explicit matrix form of such a potential,
\begin{equation}
V = \left(
\begin{matrix}
v_{0} + c b_{3} & c \left( b_{1} - ib_{2}\right) \\
c \left( b_{1} + ib_{2}\right)  & v_{0} - cb_{3}%
\end{matrix}
\right)  \! \! \: = \! \! \: \left(
\begin{matrix}
v_{0} + c b_{3} & c b_{12} e^{-i\delta_{12}}\\
c b_{12} e^{i\delta_{12}} & v_{0} - cb_{3}%
\end{matrix}
\right) \label{rft}%
\end{equation}
with writing $b_{1} + ib_{2}$ in the polar form $b_{12}e^{i\delta_{12}}$.
{}From this we can read off that, if $\delta_{12}$ is space-independent, then
this potential is phase self-transpose.

We close this Section with the simple observation -- which is easy to derive
from (\ref{ras}) -- that a conjugation (\ref{rfp}) by a ${\hat{U}}$
(\ref{rfo}) acts on $V$ like the unitary transformation of the polarization
basis (\ref{rfn}) that corresponds to some rotation around the $\sigma_{3}$
axis. This side remark will find application in Part~\ref{rac}.

\subsection{Forward transmission processes\label{rfb}}

There is an important special type of forward scattering (${\mathbf{k}}%
_{\beta}={\mathbf{k}}_{\alpha}={\mathbf{k}}$, $k=|{\mathbf{k}}|$) settings
where the scattering solution $\chi_{\alpha}$ [cf.\ (\ref{Lip1})] equals
$u_{\alpha}$ before reaching the scatterer object and is $T_{{\mathrm{f}}%
}u_{\alpha}$ afterwards, where the $2\times2$ forward transmission matrix
$T_{{\mathrm{f}}}$ is an analytic function of $V$. For example, in
M\"{o}ssbauer optics, the wave crosses a layer of width $d$ perpendicularly,
the potential $V$ is constant within the layer and is zero outside, and
\begin{equation}
T_{{\mathrm{f}}}^{} = e^{i\left( kd + \frac{d}{2k}V\right) } = g_{0}\sigma_{0}
+ gV,\label{rga}%
\end{equation}
where the exponential has been expanded using the identity
\begin{align}
e^{M}  &  = e^{\frac{1}{2}{\mathrm{tr}}M}\left( \cos \sqrt{\det N} \sigma_{0} +
\frac{\sin \sqrt{\det N}}{\sqrt{\det N}}N\right) ,\nonumber \\
N  &  = M - {\frac{1}{2}} \left( {\mathrm{tr}M}\right)  \sigma_{0}\label{rfv}%
\end{align}
for $2\times2$ matrices $M$ -- a straightforward consequence of (\ref{rde}) -- 
and the $V$ dependence of the complex multiplier scalars $g_{0}, g$ is not
displayed because only the matrix structure of $T_{{\mathrm{f}}}$ will be
relevant for the present considerations.

Indeed, from this expanded form it is easy to observe that $T_{{\mathrm{f}}}$
is  \newline-- self-transpose, \newline-- phase self-transpose,  \newline--
univectorial  \newline if and only if $V$ is
respectively  \newline-- self-transpose,  \newline-- phase self-transpose,
\newline-- univectorial.

Now, for such forward transmission situations, the task to calculate
$\left \langle \beta \left \vert T\right \vert \alpha \right \rangle $ is reduced to
determine $\left(  p_{\beta}, T_{{\mathrm{f}}} p_{\alpha}\right) $. In
accordance with this, the reciprocity theorem of the transition amplitude
simplifies to
\begin{equation}
\big ( p_{\beta}, T_{{\mathrm{f}}}^{} p_{\alpha}\big ) = \big ( p_{\overline
{\alpha}}, T_{{\mathrm{f}}}^{} p_{\overline{\beta}}\big ).\label{rfw}%
\end{equation}
Further, if $V$ is univectorial then $\delta_{12}$ [cf.\ (\ref{rft})] is
space-independent so $V$ is phase self-transpose. However, more is true for
such cases. Namely, a unitary transformation of the polarization basis
(\ref{rfn}) acts on ${\mathbf{b}}$ of a univectorial $V$, hence, $\delta_{12}$
of a space-independent ${\mathbf{b}}$ remains space-independent after any
rotation of ${\mathbf{b}}$. The consequence is that, in these forward
transmission settings, a homogeneous univectorial $V$ exhibits magnitude
reciprocity with respect to \textit{any} (essentially $2\times2$) antiunitary
operator $K$.

The question of reciprocity for transmission through more than one such layer
can be analyzed the same way as for general forward scattering in the Born
approximation (see the following Section).

\subsection{The Born approximation\label{rez}}

The importance of the (1$^{{\mathrm{st}}}$) Born approximation
[cf.\ (\ref{rcl})--(\ref{rdp})] lies in its validity for many situations,
\textit{e.g.,} for scattering on thin enough layers. In parallel, from the
aspect of reciprocity [cf.\ (\ref{rcx}) and (\ref{rey})], it plays a special
role as the violation of reciprocity or magnitude reciprocity may vanish in
the Born approximation. Let us now consider its details for two polarization
degrees of freedom.

In the Born approximation, the scattering amplitudes (\ref{scatamplT+-})
become linear in the incoming polarization:
\begin{equation}
f({\mathbf{k}}_{\beta},\alpha)_{j}^{}\approx f({\mathbf{k}}_{\beta
},{\mathbf{k}}_{\alpha})_{jk}^{}p_{k}^{}\label{ret}%
\end{equation}
for each of the four types of scattering amplitude. Correspondingly, the
transition amplitude simplifies to the two-component scalar product
\begin{equation}
\left(  p_{\beta}, \left[ \mbox{\Large$\int$} e^{i\left( {\mathbf{k}}_{\alpha}
-{\mathbf{k}}_{\beta}\right) {\mathbf{r}}}V({\mathbf{r}}) {\mathrm{d}}%
^{3}{\mathbf{r}} \right] p_{\alpha}\right) \label{rfx}%
\end{equation}

In case of forward scattering, it reduces to
\begin{equation}
\left(  p_{\beta}, \left[ \mbox{\Large$\int$} V({\mathbf{r}}) {\mathrm{d}}%
^{3}{\mathbf{r}} \right] p_{\alpha}\right) \label{rfy}%
\end{equation}
If $V$ is piecewise constant, $V_{l}$ on a spatial region of volume
${\mathcal{V}}_{l}$, with $l = 1, \ldots, n$ (\textit{e.g.,} a sample
constituted by $n$ homogeneous layers) then (\ref{rfy}) gives
\begin{equation}
\left(  p_{\beta}, \left( \mbox{\large$\sum$}_{l=1}^{n}{\mathcal{V}}_{l}%
V_{l}\right)  p_{\alpha}\rule{0em}{2.ex}\right) .\label{rfz}%
\end{equation}
If, further, each $V_{l}$ is univectorial such that ${\mathbf{v}}%
_{l}=c{\mathbf{b}}_{l}$ with $l$-independent $c$ then the sum is a
univectorial matrix with Poincar\'{e} three-vector $\,c\sum_{l} {\mathcal{V}%
}_{l}{\mathbf{b}}_{l}\,$ and, consequently, we have magnitude reciprocity in
any orthonormal polarization basis [see (\ref{rfa})].

\section{Examples and applications\label{rac}}

The general treatment of reciprocity (and nonreciprocity) given in
Part~\ref{raa} has been investigated in large detail for two-component wave
functions in Part~\ref{rab} because of the many related important applications
in elastic scattering of photons and neutrons.

We recall the result given in Sect.~\ref{rez} that nonreciprocal forward
scattering disappears under the conditions of the first Born approximation,
namely, in case of weak scattering. The recoil-less nuclear resonance forward
scattering of photons, known as M\"{o}ssbauer scattering, can realize the case
of strong scattering. Indeed, soon after the discovery of the M\"{o}ssbauer
effect the so-called blackness effects were reported and identified as the
result of multiple scattering. In the theoretical description of the
M\"{o}ssbauer scattering given by \textcite{Blume68}, the ensemble of nuclei
and electrons, as scattering centers, represent an anisotropic and absorbent
optical medium, which is described by a $2\times2$ complex index of refraction
$n$ \cite{Lax51} corresponding to the two possible independent states of
polarization \cite{Blume68}. The index of refraction is related simply to the
coherent forward-scattering amplitude $f$ and to the number of scattering
centers per unit volume $N$, in the form $\, n=\sigma_{0}+\left(  2\pi
N/k^{2}\right)  f\, $, where $k$ is the wave number in vacuum \cite{Lax51}.
For photons, $f$ is the sum of the electronic and nuclear scattering
amplitudes, $f=f_{{\mathrm{e}}}+f_{{\mathrm{n}}}$ \cite{Hannon85b} and, for
neutrons, it is the sum of the nuclear and magnetic scattering lengths,
$f=f_{{\mathrm{nuc}}}+f_{{\mathrm{magn}}}$. In each homogeneous part around
position ${\mathbf{r}}$, an index of refraction $n\left(  {\mathbf{r}}\right)
$ can be defined, which can be interpreted as an optical potential in a wave
equation
\begin{equation}
\left[  \Delta+k^{2}I\right]  \psi \left(  {\mathbf{r}}\right)  =V\left(
{\mathbf{r}}\right)  \psi \left(  {\mathbf{r}}\right) \label{mosscat}%
\end{equation}
for a two-component wave function $\psi$ via the relation $V\left(
{\mathbf{r}}\right)  =2k^{2}\left[  \sigma_{0}-n\left(  {\mathbf{r}}\right)
\right]  $ \cite{BornWolf}. Here, the two components describe the two
polarizations of the photon field, but actually a same type of equation can be
used for neutrons as well. Indeed, based on the equation (\ref{mosscat}), the
elastic scattering of slow neutrons and of X-rays on stratified media have a
common description \cite{Deak01}, which also covers M\"{o}ssbauer scattering
\cite{Deak96} and diffuse scattering. The common scattering theory is that of
the (time dependent) Schr\"{o}dinger equation whose stationary scattering
states satisfy (\ref{mosscat}).

We note that the elements of the matrix $n$ for both slow neutrons and X-rays
differ only slightly (typically $10^{-5}$) from that of the $2\times2$ unit
matrix $\sigma_{0}$, which difference can be three orders of magnitude
greater, typically $10^{-2}$, in the case of M\"{o}ssbauer scattering at the
resonance energies. This difference is the basis of speaking about stronger
scattering in case of the M\"{o}ssbauer medium. In the so-defined optical medium,
namely, in the M\"{o}ssbauer medium, the magneto-optic Faraday effect was
identified \cite{Blume68}, which is often cited as a property of
nonreciprocal media \cite{Kamal2011}. Both aspects, the presence of strong
scattering and the magneto-optic effects, recommend using M\"{o}ssbauer medium
as a model system for studying reciprocity.

In the following two Sections, some examples of M\"{o}ssbauer forward
scattering on pure $\alpha-^{57}\mathrm{Fe}$ absorbers follow, to demonstrate
reciprocal and nonreciprocal (in other words, reciprocity violating) cases.
All the considered iron foils had the thickness of $\, 4\, \mathrm{\mu m}$,
and the spectra were simulated by the computer program EFFI \cite{Spiering00},
which reproduces experimental results to high preciseness. As we will see, the
classification of situations into reciprocal and nonreciprocal provides
unexpected outcomes sometimes.


\subsection{When nonreciprocity is expected but reciprocity appears, instead
\label{circmossba}}

It was already mentioned that, in the literature, magneto-optic media is often
cited as nonreciprocal media. In an early work on the field
\cite{Frauenfelder1962}, it was pointed out that the sign of the $\mathbf{k}%
$--parallel component of the magnetization can be determined using circularly
polarized radiation. Simply put, the situations when the (homogeneous)
magnetization $\mathbf{B}_{\mathrm{Hf}}$ is parallel to the momentum vector
$\mathbf{k}$ can be distinguished from the antiparallel case.

\begin{figure}[p]
\resizebox{.6\columnwidth}{!}{\includegraphics[clip]{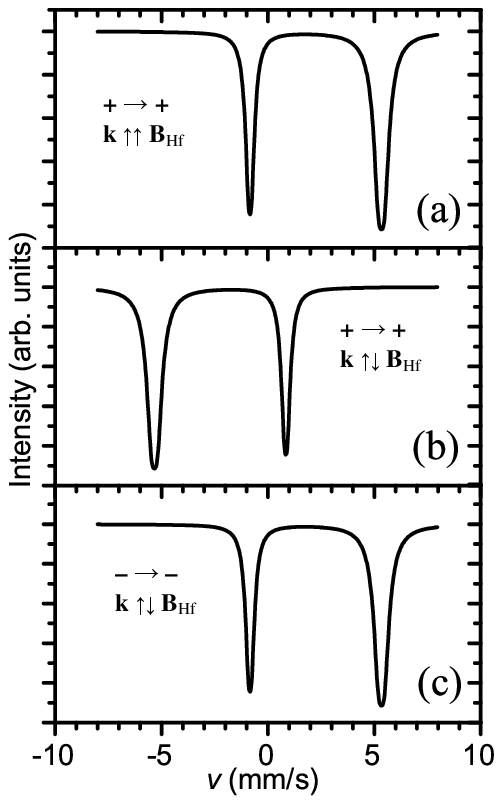}} 
\caption{Simulated
M\"{o}ssbauer forward scattering spectra on $\mathrm{\alpha}-^{57}\mathrm{Fe}$
foil of thickness of $4\  \mathrm{\mu m}$ using the polarizer-analyzer setup
measured for the cases: (a) incident right circular polarized photons
scattered to the same polarization ($+\longrightarrow+$ scattering) and the
hyperfine magnetic field $\mathbf{B}_{\mathrm{Hf}}$ being parallel to the
momentum vector $\mathbf{k}$\textbf{, }(b) incident right circular polarized
photons scattered to the same polarization ($+\longrightarrow+$ scattering)
and the hyperfine magnetic field $\mathbf{B}_{\mathrm{Hf}}$ being antiparallel
to the momentum vector $\mathbf{k}$, and (c) incident left circular polarized
photons scattered to the same polarization ($-\longrightarrow-$ scattering)
and the hyperfine magnetic field $\mathbf{B}_{\mathrm{Hf}}$ being antiparallel
to the momentum vector $\mathbf{k}$.}%
\label{fig2}%
\end{figure}

At first sight, it seems that the here-defined arrangement is nonreciprocal.
Indeed, the change of the magnetization from parallel to antiparallel case can
be asked whether it can be a `reciprocal-plus-rotational' transformation,
discussed in Sect.~\ref{Rotavsreci}, and by (\ref{rev}), the answer seems to
be negative. In fact, a mere change of the direction of magnetization does
modify the M\"{o}ssbauer spectra, as shown in Figs.~\ref{fig2} (a) and
\ref{fig2} (b).

However, taking into account the incoming and outgoing polarizations as
well, if we prescribe a simultaneous change of the incoming and outgoing
-- say, right circular -- polarizations to opposite, then the spectrum
remains invariant, as applying (\ref{rev}) gives, and comparing
Figs.~\ref{fig2} (a) and \ref{fig2} (c) justifies. Consequently, the
magneto-optic medium discussed here itself is reciprocal (!), and this
property either remains hidden or is revealed depending on how one
chooses the polarizations for the processes.

This reciprocal property can be easily understood considering the scattering
potential in the case of longitudinal Zeeman effect, in which case the
potential is diagonal on the circular basis, fulfilling thus the requirement
of self-transposeness, which is a manifest form of the reciprocity condition.

\subsection{Real nonreciprocity in M\"{o}ssbauer scattering
\label{nonrecmossba}}

Unlike in the previous example, where circularly polarized photons scattered
on a \textit{single} $\alpha-^{57}\mathrm{Fe}$ M\"{o}ssbauer absorber, next
considered is the case of scattering of linearly polarized photons -- the
scattering from incident TE (transversal electric, in other words, $\sigma
$--polarization) to the same type of outgoing polarization -- on \textit{two}
$\alpha-^{57}\mathrm{Fe}$ M\"{o}ssbauer absorbers. The case of reciprocity for
more than one scatterer was considered in section \ref{rad} with the
conclusion that piecewise univectorial potentials admit a reciprocity operator
if and only if the corresponding real vectors ${\mathbf{b}}_{l}$ of the
piecewise constant potential values $V_{l}$ are within a common plane. Since
the potential in case of magnetic hyperfine interaction is of univectorial
kind (see the next paragraph), in the two-scatterers case ${\mathbf{b}}_{l}$
($l=1,2$) are necessarily within a common plane, hence, there are always
reciprocal situations. We note that this is not true for the case of three
scatterers with three linearly independent vectors, as three such vectors do
not lie in a common plane.

The polarization-dependent part of the scattering amplitude for M\"{o}ssbauer
transition in case of $^{57}$Fe can be given following Eq.~5 of
Ref.~\cite{Hannon69} and using the deduction of \textcite{Rose57}. One finds
that the scattering amplitude $f$, consequently the optical potential $V$, is
written as a product of a complex number (the energy-dependent complex
Lorentzian) and a self-adjoint matrix (the polarization-dependent part of $f$)
and is, therefore, univectorial. The corresponding real vector $\mathbf{b}$
of the +1 transition is, after some simple algebra, found to be
\begin{equation}
\mathbf{b}=\left(
\begin{array}
[c]{c}%
-\frac{1}{2}\sin^{2}\theta \sin2\varphi \\
-2\cos \theta \\
\sin^{2}\theta \cos2\varphi
\end{array}
\right)  ,\label{V_poincare}%
\end{equation}
where a linear basis for the polarization is used with $\theta$ and $\varphi$
being the polar angles describing the direction of the hyperfine magnetic
field in the laboratory system, which is selected so that the $z$ axis points
towards the propagation of the radiation and the two other axes are arbitrary
orthogonal directions.

\begin{figure}[p]
\resizebox{.6\columnwidth}{!}{\includegraphics[clip]{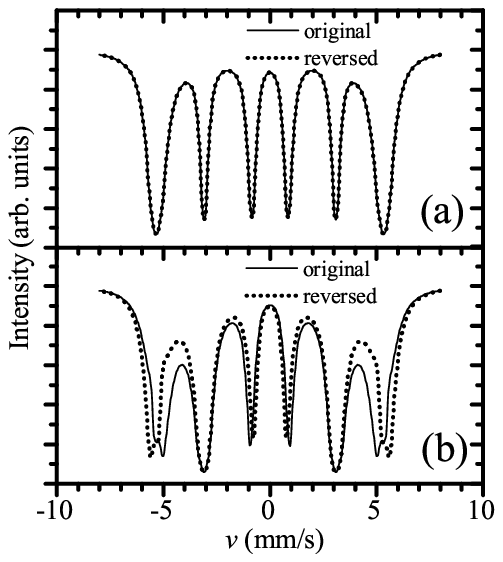}} 
\caption{Simulated M\"{o}ssbauer forward scattering spectra on two
$\mathrm{\alpha}-^{57}\mathrm{Fe}$ foils of thicknesses of $4\  \mathrm{\mu m}$
using the polarizer-analyzer setup for incident $\sigma$ (\textit{viz}.
transversal electric) linearly polarized photons scattered to the same
polarization ($\sigma \longrightarrow \sigma$ scattering) for the cases the
hyperfine magnetic fields $\mathbf{B}_{\mathrm{Hf,1}}$ and $\mathbf{B}%
_{\mathrm{Hf,2}}$ in the foils pointing to the directions given by the polar
angles (a) $\theta_{1}=90^{\circ}$, $\varphi_{1}=90^{\circ}$ and $\theta
_{2}=135^{\circ}$, $\varphi_{2}=0^{\circ}$, and (b) $\theta_{1}=90^{\circ}$,
$\varphi_{1}=45^{\circ}$ and $\theta_{2}=135^{\circ}$, $\varphi_{2}=0^{\circ}%
$. In both parts, (a) and (b), solid lines denote the original and dashed
lines the reversed (source-detector exchanged) situations. }%
\label{fig3}
\end{figure}

Let us consider the arrangement of two samples, defined by the polar angles of
the hyperfine fields $\theta_{1}=90^{\circ}$, $\varphi_{1}=90^{\circ}$ and
$\theta_{2}=135^{\circ}$, $\varphi_{2}=0^{\circ}$
. The vectors $\mathbf{b}_{1}$, $\mathbf{b}_{2}$ of the corresponding two
potentials of the two samples read $\mathbf{b}_{1}=\left(  0,0,-1\right)  $
and $\mathbf{b}_{2}=\left(  0,\sqrt{2},1/2\right)  $. The reciprocal
arrangement can be defined according to the geometrical criterion of
Sect.~\ref{rad}, namely, via the unitary transformation that expresses a
rotation around the intersection of the $\sigma_{1}$--$\sigma_{3}$ plane and
the plane spanned by the vectors $\mathbf{b}_{1}$ and $\mathbf{b}_{2}$, and
brings the former plane to the latter. This arrangement has the speciality
that the vector $\mathbf{b}_{1}$ also lies in the $\sigma_{1}$--$\sigma_{3}$
plane and, as a consequence, it is parallel to the rotational axis, which can
only be the $\sigma_{3}$ axis in this case. In Sect.~\ref{rfa} it was
mentioned that any rotation around the $\sigma_{3}$ axis causes magnitude
reciprocity. Indeed, Fig.~\ref{fig3} (a) shows that the corresponding
simulated M\"{o}ssbauer spectra are identical for the normal and reciprocal cases.

Next, we present an arrangement, a slight modification of the previous
example, where real observable nonreciprocity appears. The modification is
only that we rotate the first foil by $45^{\circ}$ around the axis being
parallel to the direction of the wave propagation $\mathbf{k}$, resulting in
the polar angles of $\theta_{1}=90^{\circ}$, $\varphi_{1}=45^{\circ}$ and
$\theta_{2}=135^{\circ}$, $\varphi_{2}=0^{\circ}$
. Repeating the previous procedure, the $\mathbf{b}$ vectors of the
corresponding potential read $\mathbf{b}_{1}=\left(  -1,0,0\right)  $ and
$\mathbf{b}_{2}=\left(  0,\sqrt{2},1/2\right) $. In this case, $\mathbf{b}%
_{1}$ is a vector both of the $\sigma_{1}$--$\sigma_{3}$ plane and of the
plane spanned by the vectors $\mathbf{b}_{1}$ and $\mathbf{b}_{2}$, therefore,
the axis of the rotation $U$ is the $\sigma_{1}$ axis. Unlike in the previous
example, such a rotation does not cause magnitude reciprocity, thus observable
nonreciprocity is expected. Indeed, the simulations shown in Fig.~\ref{fig3}
(b) confirm the nonreciprocal property of the arrangement. Concentrating on
the first M\"{o}ssbauer lines (left ones) of the normal and reciprocal
arrangements, one can conclude that the type of asymmetry caused by multiple
scattering appears on the opposite side of the lines in the normal and the
reciprocal arrangements. This feature fades away for thin layers (thicknesses
less than $1\  \mathrm{\mu m}$), where the 1$^{\mathrm{st}}$ Born approximation
becomes valid and, according to Sect.~\ref{rez}, nonreciprocity disappears.

We note that the here-presented thought experiment uses the so-called
polarizer-analyzer setup, therefore, if using radioactive source, one would
lose most part of the intensity. A similar experiment is, however, feasible by
using synchrotron radiation source, where the beam is collimated and the
polarizer-analyzer setup means no problem. The arrangement presented in this
example is interesting because a simple rotation by $45^{\circ}$ switches the
system from magnitude reciprocal to nonreciprocal.

\subsection{On the symmetry of diffuse $\omega$-scans}

\begin{figure}[p]
\resizebox{.6\columnwidth}{!}{\includegraphics[clip]{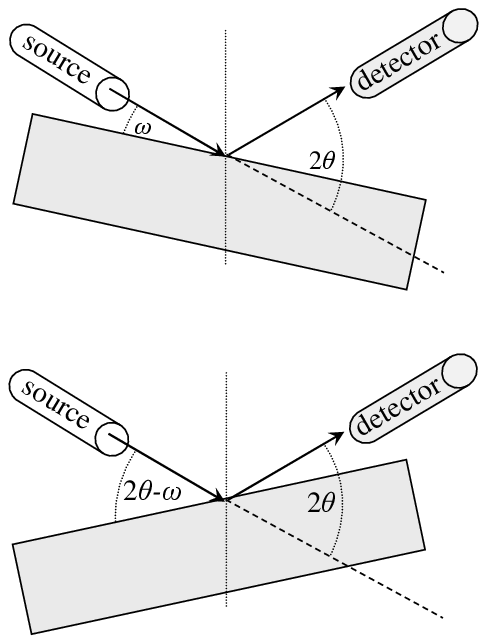}} 
\caption{Geometrical arrangement of the diffuse $\omega$-scan scattering
experiments with $2\theta$ being the angle of the incoming and outgoing waves,
and $\omega$ being the angle of the incident wave and the surface of the
stratified media. The lower part shows the setup obtained by a rotation by
$\phi=2\left(  \theta-\omega \right)  $ around the axis being perpendicular to
the scattering plane. }%
\label{fig4}
\end{figure}

For studying lateral inhomogeneities -- structural roughness, magnetic domains,
etc. -- in stratified media, diffuse scattering, \textit{i.e.,} off-specular
neutron \cite{Felcher93}, soft-X-ray resonant magnetic diffuse scattering
\cite{Hase00} and M\"{o}ssbauer reflectometry \cite{Nagy02a}, is applied. A
possible experimental realization of the off-specular reflectometry is the
so-called `$\omega$-scan' geometry, where the detector position is set to
$2\theta$ and the sample orientation $\omega$ on the goniometer is varied with
the sample normal remaining in the scattering plane (see Fig.~\ref{fig4}). It
is straightforward to see in this setup that the incoming and outgoing waves
make an angle of $\omega$ and $2\theta-\omega$ with the surface of the
stratified media, respectively. In the '$\omega$-scan' experiment, the
scattered intensity $J$\ as a function of $\omega$ is detected. In the special
case $\omega=\theta$, one detects the specular radiation. A well-known
property of the '$\omega$-scan' intensity function is its symmetricity with
respect to the specular position, \textit{i.e.,} $J\left(  \omega \right)
=J\left(  2\theta-\omega \right) $, which is explained in the
literature as a straightforward consequence of the reciprocity theorem, which
is often confused with time reversal symmetry \cite{Chernov2005}. Let us now
investigate how our reciprocity formulae can be utilized for this symmetricity property.

What we ask is whether the sample at position $\omega$, \textit{i.e.,}
potential $V$, is a reciprocal partner of the sample at position
$2\theta-\omega$, \textit{i.e.,} of potential $V_{\phi}=R_{\phi}^{}VR_{\phi
}^{-1}$, where $R_{\phi}$ is the Hilbert space representation of the rotation
by $\phi:=(2\theta-\omega)-\omega= 2\left(  \theta-\omega \right)  $ that
connects the two positions. Since the incoming and outgoing momenta are the
same for the two arrangements, we must consider reciprocity combined with a
rotation as done in Sect.~\ref{Rotavsreci}, at Eq.~(\ref{rgb}), which we
repeat here for convenience:
\begin{align}
\hskip -.7em \big \langle {\mathbf{k}}_{\beta},p_{\beta}%
\big \vert T\big \vert {\mathbf{k}}_{\alpha},p_{\alpha}\big \rangle  &
=\big \langle {\mathbf{k}}_{\beta},U_{R }Up_{\alpha}^{*}\big \vert {\overline
{T}}_{R }\big \vert {\mathbf{k}}_{\alpha},U_{R } Up_{\beta}^{*}%
\big \rangle ,\label{rgc}%
\end{align}
where the rhs describes a scattering on $R \overline{V}R ^{-1}=R
UV^{{\mathsf{T}}}U^{-1}R ^{-1}$, with $R$ being the Hilbert space
representation of the rotation performing ${\mathbf{k}}_{\alpha}%
\leftrightarrow-{\mathbf{k}}_{\beta}$.

The question is whether this, the scattering amplitude at position $\omega$,
can coincide with the one at position $2\theta-\omega$,
\begin{align}
\big \langle {\mathbf{k}}_{\beta},p_{2}\big \vert T_{\phi}%
\big \vert {\mathbf{k}}_{\alpha},p_{1}\big \rangle ,\label{rgd}%
\end{align}
with some appropriate polarizations $p_{1}$, $p_{2}$. To ensure this, on one
hand we require $V_{\phi} = R \overline{V}R ^{-1}$, which can be expanded and
rearranged as
\begin{align}
R_{\phi}^{}VR_{\phi}^{-1}  &  = R UV^{{\mathsf{T}}}U^{-1}R ^{-1},\nonumber \\
R ^{-1}R_{\phi}^{}VR_{\phi}^{-1}R  &  = UV^{{\mathsf{T}}}U^{-1},\nonumber \\
R_{{\mathbf{m}},\pi} ^{}VR_{{\mathbf{m}},\pi}^{-1}  &  = UV^{{\mathsf{T}}%
}U^{-1};\label{rge}%
\end{align}
here, in the last line, we recognized that the combination of the two
rotations $R=R^{-1}$ and $R_{\phi}$ is the rotation $\, R_{{\mathbf{m}},\pi
}^{}=R_{{\mathbf{m}},\pi} ^{-1}\, $ of the sample in position $\omega$ around
its normal ${\mathbf{m}}$ by $\pi \equiv180^{\circ}$. Expressed in the
Poincar\'{e} vector
description, condition (\ref{rge}) says
\begin{align}
v_{0}\left( {\mathbf{O}}_{{\mathbf{m}},\pi}{\mathbf{r}}\right)   &  =
v_{0}\left( {\mathbf{r}}\right) \label{rgf}\\
{\mathbf{O}}_{{\mathbf{m}},\pi}{\mathbf{v}}\left(  {\mathbf{O}}_{{\mathbf{m}%
},\pi}{\mathbf{r}}\right)   &  = {\mathbf{O}}_{U}{\mathbf{P}}_{13}{\mathbf{v}}
\left( {\mathbf{r}}\right) ,\label{rgh}%
\end{align}
where ${\mathbf{O}}_{U}$ is the rotation belonging to $U$ by (\ref{ras}). We
note that the ${\mathbf{r}}$-dependence aspect may be successfully treated for
lateral inhomogeneities via the DWBA approximation. In parallel, (\ref{rgh})
can be evaluated analogously to our previous analyses.

The other requirement is that the polarizations also agree. Namely, if $V$
satisfies (\ref{rgf})--(\ref{rgh}) with some $U$ then one needs
\begin{align}
&  & p_{1}  &  = U_{R}Up_{\beta}^{*}, & p_{2}  &  = U_{R}Up_{\alpha}^{*}. &  &
\label{rgg}%
\end{align}
Examples for such polarization settings were considered in
Sect.~\ref{Rotavsreci}.

As for simple examples, the simplest one is that of a polarization independent
potential, $V({\mathbf{r}}) = v_{0}({\mathbf{r}}) \sigma_{0}$. Even for such
potentials, (\ref{rgf}) prescribes a nontrivial condition so even polarization
independent potentials must obey such a rotation invariance so as to exhibit
the symmetry of the $\omega$-scan spectrum.

The second example is neutron scattering on a sample of one layer with depth
independent magnetic field, which is a homogeneous ${\mathbf{B}}_{\text{l}}$
in the left half of the sample and a homogeneous ${\mathbf{B}}_{\text{r}}$ in
the right half (let the shape of the sample be left-right reflection
invariant, with respect to a plane orthogonal to the layer). Then (\ref{rgh})
requires
\begin{equation}
{\mathbf{B}}_{\text{l}} = {\mathbf{O}}_{{\mathbf{m}},\pi}{\mathbf{O}}%
_{U}{\mathbf{P}}_{13}{\mathbf{B}}_{\text{r}}.\label{rgl}%
\end{equation}
The combined transformation ${\mathbf{O}}_{{\mathbf{m}},\pi}{\mathbf{O}}%
_{U}{\mathbf{P}}_{13}$ is an orthogonal, thus length-preserving, one. If
$|{\mathbf{B}}_{\text{l}}| = |{\mathbf{B}}_{\text{r}}|$ then one can find such
a definition of the $\sigma_{1}$ and $\sigma_{3}$ directions -- the $x$ and $z$
directions -- that (\ref{rgl}) is satisfied. On the other hand, if
$|{\mathbf{B}}_{\text{l}}| \not = |{\mathbf{B}}_{\text{r}}|$ then it is
impossible to fulfill this requirement and the spectrum of the $\omega$-scan
cannot be symmetric.

\section{Conclusion}

As demonstrated in the general-level discussion, reciprocity is the property
of a system -- describing linear wavelike propagation -- when it is connected
with the adjoint system by an antiunitary operator. If the reciprocity
property is fulfilled via such a reciprocity operator $K$ then a reciprocity
theorem holds, which expresses the equality of any scattering amplitude
$\langle \beta|T|\alpha \rangle$ to another scattering amplitude, namely, to
$\langle K\alpha|T|K\beta \rangle$. This left-right interchange of incoming and
outgoing states gives, in important experimental applications, that a
scattering amplitude is related to another one where the source and the
detector are interchanged.

Remarkably, reciprocity can hold for non-selfadjoint systems (systems with
absorption), too. This immediately distinguishes reciprocity from time
reversal invariance. Waves with spin/polarization degree of freedom
demonstrate that reciprocity also differs from rotational invariance:
Rotations are unable to map an incoming polarization degree of freedom to an
outgoing one, nor an outgoing polarization to an incoming one. The
above-presented calculations show in detail the relationship of reciprocity to
time reversal as well as how rotation can be combined with reciprocity to
obtain a version of the reciprocity theorem that is especially suitable for
scattering experiments.

To find reciprocity operators for a given system is a delicate problem, which
is solved here for an important class of physical situations, which cover
applications in neutron and photon scattering on multilayer structures.
Reciprocity violation is also quantified, and the results are illustrated and
applied on examples, chosen from the area of M\"ossbauer scattering, where
reciprocity is fulfilled for certain processes and is immensely violated for
some others (scattering amplitudes differing remarkably).

The relationship established here to a recently developing area of mathematics
is expected to give an impetus to finding reciprocity operators to more
physical systems, resulting in valuable applications.

\begin{acknowledgments}
This work was partly supported by the Hungarian Scientific Research Fund
(OTKA), the National Office for Research and Technology of Hungary (NKTH), and
the European Research Council under contract numbers K81161, NAP-Veneus'08,
and StG-259709, respectively. Our gratitude goes to \textit{Hartmut Spiering}
(Johannes Gutenberg University, Mainz) for the software development. 
\end{acknowledgments}

\bibliographystyle{apsrev}

\end{document}